\documentclass[twocolumn, times, trackchanges]{aastex63}

\usepackage{enumitem}
\usepackage{amsmath}
\usepackage{longtable}
\usepackage{latexsym}
\usepackage{graphicx}
\usepackage{epsf}
\usepackage{CJK}

\renewcommand{\vec}[1]{\mathbf{#1}}

\received{June 21, 2021}
\revised{January 18, 2022}
\accepted{January 31, 2022}

\submitjournal{AJ}

\shorttitle{Stellar Flybys in Sco-Cen}
\shortauthors{Ma, De Rosa, Kalas}

\graphicspath{{./}{figures/}}

\begin{document}
\begin{CJK*}{UTF8}{gbsn}
\title{Search for Stellar Flybys in the Sco-Cen OB Association with the Gaia DR2}

\author[0000-0002-0463-9528]{Yilun Ma (马逸伦)}
\affiliation{Department of Astronomy, University of California, Berkeley, CA 94720, USA}

\author[0000-0002-4918-0247]{Robert J. De Rosa}
\affiliation{European Southern Observatory, Alonso de C\'{o}rdova 3107, Vitacura, Santiago, Chile}

\author[0000-0002-6221-5360]{Paul Kalas}
\affiliation{Department of Astronomy, University of California, Berkeley, CA 94720, USA}
\affiliation{SETI Institute, Carl Sagan Center, 189 Bernardo Ave., Mountain View CA 94043, USA}
\affiliation{Institute of Astrophysics, FORTH, GR-71110 Heraklion, Greece}

\begin{abstract}
High-contrast imaging studies of debris disks have revealed a significant diversity in their morphologies, including large-scale asymmetries. Theories involving stellar flybys, an external source of gravitational disturbance, have offered a plausible explanation for the origin of these morphological variations. Our study is an experiment to gain empirical evidence that has been lacking from such theories. We explore this paradigm by using astrometric and radial velocity measurements from the \textit{Gaia} DR2 and ground-based observations to trace the trajectories of 625 stars in the Sco-Cen OB Association from 5 Myr in the past to 2 Myr in the future.
We identified 119 stars that had at least one past flyby event occurring within one Hill radius, and 23 of these experienced flybys within 0.5 Hill radius.
We found no evidence of a significant correlation between the presence of flyby events and infrared excess detections, although the sample is not uniformly sensitive to infrared excess emission. Ten stars that had past flyby events host resolved circumstellar disks that appear relatively symmetric in the existing data except for the circumbinary disk surrounding HD 106906. We determined the trajectory and relative velocity of each of these flyby events, and compared these to the geometry of the spatially-resolved disks. Future work is needed to measure the kinematics of lower mass stars and to improve sensitivity to circumstellar disks for the entire sample. 

\end{abstract}


\section{Introduction} \label{sec:intro}
Our planetary system is mostly flat and symmetric with the terrestrial and giant planets orbiting nearly in the same plane with low eccentricities. However, highly eccentric orbits and misalignments in extrasolar planetary systems (e.g., high mutual inclinations, disk-planet misalignments, disk asymmetries, etc.) are not unusual (e.g., \citealt{kalas95a,tamuz08a,triaud10a,kennedy12a,derosa20a}). One mechanism to explain high eccentricities is planet-planet scattering, a form of internal dynamical excitation (e.g., \citealt{1996Sci...274..954R, raymond10a}). How the scattering starts out in the first place can be attributed to planet migration in gaseous circumstellar disks, thus disrupting the primordial orbital configuration of the system \citep{2012ARA&A..50..211K}.

\par However, external stellar and sub-stellar perturbers can also reshape the architectures of planetary systems. Such perturbers can be bound to the system in cases of stellar multiplicity or distant planet-mass objects, or unbound in the case of stellar or sub-stellar flybys. An example of the bound stellar case is the RW Aur system.  \cite{2006A&A...452..897C} and \cite{2018ApJ...859..150R} suggested that periastron passages between RW Aur A and B could explain the asymmetric features of the RW Aur B disk and the trailing material from the RW Aur A disk. An example of the bound sub-stellar case is the 11 M$_J$ mass exoplanet HD 106906 b which orbits exterior to the distorted circumbinary debris disk  \citep{kalas_2015} and which could explain the disk's asymmetric morphology \citep{nesvold17a, rodet17a, 2021AJ....161...22N}. If the planet originally formed in the disk and evolved outward due to interior instabilities, unbound stellar flybys may have raised its periastron distance away from the central planetary region \citep{rodet17a, derosa_kalas}. Thus the HD 106906 system has evidence for both internal and external perturbations at play. 

Stellar flyby events have occurred during our own solar system's history as well. The existence of dynamically new, long-period comets entering the inner solar system at random inclinations is consistent with a scenario where primitive icy bodies were ejected to large aphelia by the giant planets, had their perihelia raised and inclinations randomized by perturbations from passing stars, and approach the Sun again at later epochs when their perihelia are decreased by other passing stars \citep{oort50a, hills81a, duncan87a}.  For example, \cite{2015ApJ...800L..17M} found that the low-mass binary WISE J072003.20084651.2---also known as ``Scholz's star"---had a flyby with our Sun at 0.25 pc roughly 70,000 years ago. This is the closest flyby event with our solar system discovered to date. Using $Gaia$ DR2 data, \citet{bailer-jones18a} calculate that Gl 710 (HIP 89825) will come within 0.09 pc of the Sun $\sim$1.2 Myr in the future and estimate that roughly 20 stars per Myr pass within 1 pc of the Sun.

Alternately, long-period comets may have been captured at early times (e.g., age $<$10 Myr) when the Sun still resided in its a birth cluster and small bodies were ejected from their original formation sites around other stars due to flybys \citep{levison10a}. The high mutual inclination and eccentricity of certain Kuiper Belt objects may be signatures of past stellar flybys of the proto-Sun with other members of the Sun's natal cluster after the gas giants had formed \citep{2000ApJ...528..351I,kenyon04a, 2005Icar..177..246K, pfalzner18a}.

Theoretical work has generally explored how the architectures of exoplanetary systems could be modified in a young cluster environment by flybys.  At the earliest epochs the structure of protoplanetary disks may be altered, such as by having disk outer radii truncated \citep{larwood97a, 2014MNRAS.441.2094R, 2018MNRAS.475.2314W, cuello19a}. Close stellar encounters can also create spiral structure, asymmetric rings, and vertically disturbed populations of particles that may account for observed debris disk asymmetries \citep{larwood01a,reche09a,lestrade11a, cuello20a}.
Encounters between a planetary system and a flyby star could excite the eccentricity of the outer planets and transfer the excitation of eccentricity inward on a longer timescale \citep{2004AJ....128..869Z}.  Flybys could also completely eject outer planets from the system and leave the surviving planetary system unstable  \citep{hills84a, 2011MNRAS.411..859M, parker12a}.

\par In this paper, we extend the \cite{derosa_kalas} study of the HD 106906 system to the entire Sco-Cen OB association, identifying the possible stellar flybys among its members using a combination of astrometry and radial velocities from \textit{Gaia} DR2 \citep{gaia2018} and from various other literature sources. In Section \ref{sec:sample}, we describe the characteristic of the samples used in this study, such as mass estimation and sample completeness. In Section \ref{sec:flyby_method}, we present the procedure used to identify stellar flybys in the samples. The results that comes out of those methods are presented in Sections \ref{sec:flyby_statistics} and  \ref{sec:field_stars}. In Section \ref{sec:ir_excess}, we discuss the correlation between the presence of infrared excess (IR) and stellar close approaches. Section \ref{sec:interaction} then focuses on close approaches involving stars hosting resolved debris disk and presents the flyby geometry of those with determined disk orientations. Section \ref{sec:usco_encounters} discusses potential encounters identified in the Upper Scorpius subgroup via simulating precise RV measurements. Lastly, Section \ref{sec:conclusion} summarizes the findings and discusses potential future development.

\section{Sample Properties}
\label{sec:sample}
\subsection{Sample Selection}\label{sec:sample_selection}
\begin{figure*}
    \includegraphics[width=\textwidth]{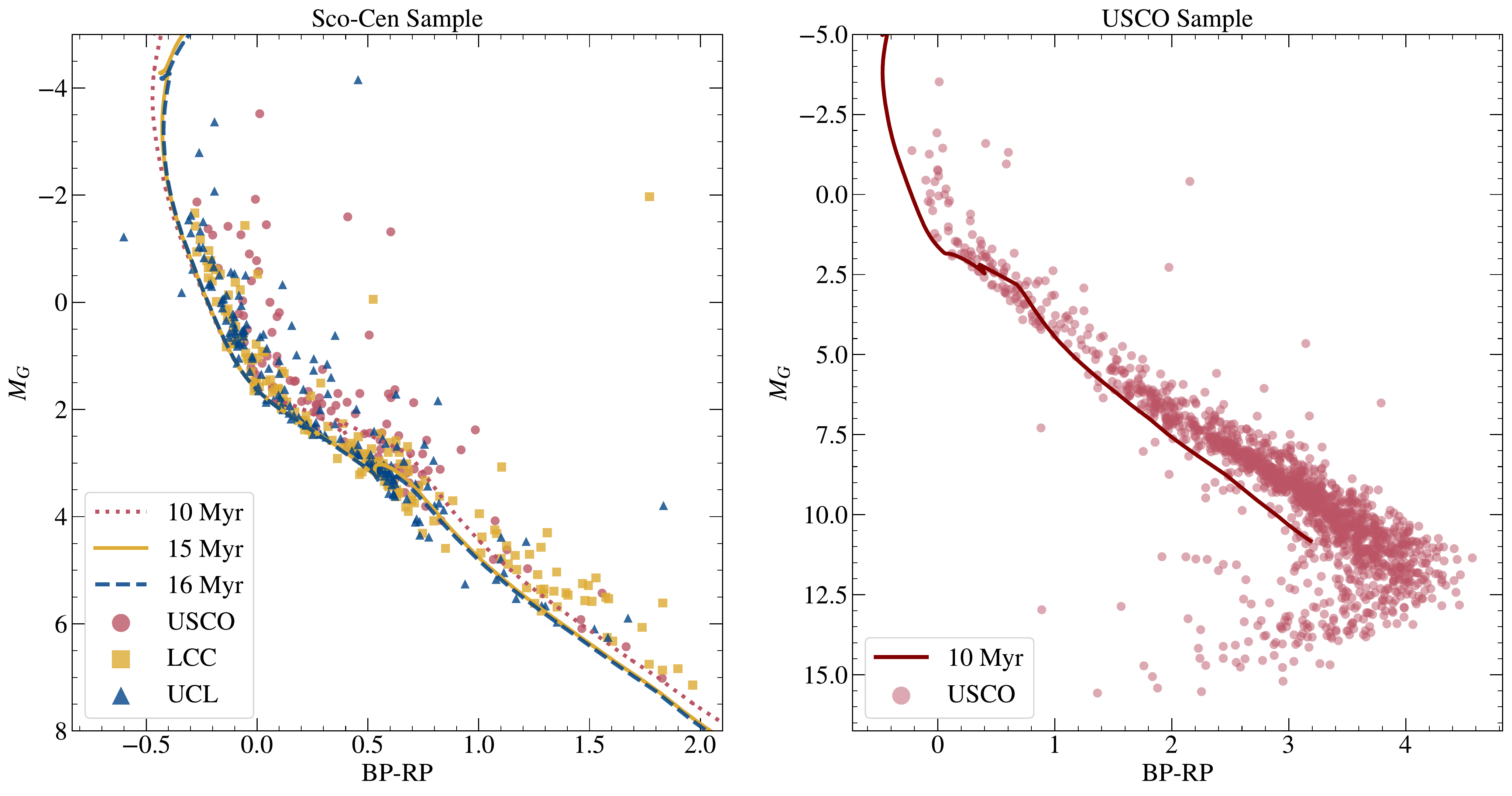}
    \caption{Color-magnitude diagram of the two samples created using optical photometry from \textit{Gaia} DR2. Some stars that are not on the main sequence are not plotted. Isochrones from MIST at the ages of the three subgroups are also plotted for comparison. No extinction correction is applied to either the sample or the isochrones.}
    \label{fig:CMD}
\end{figure*}

\begin{figure}[ht]
    \includegraphics[width=\columnwidth]{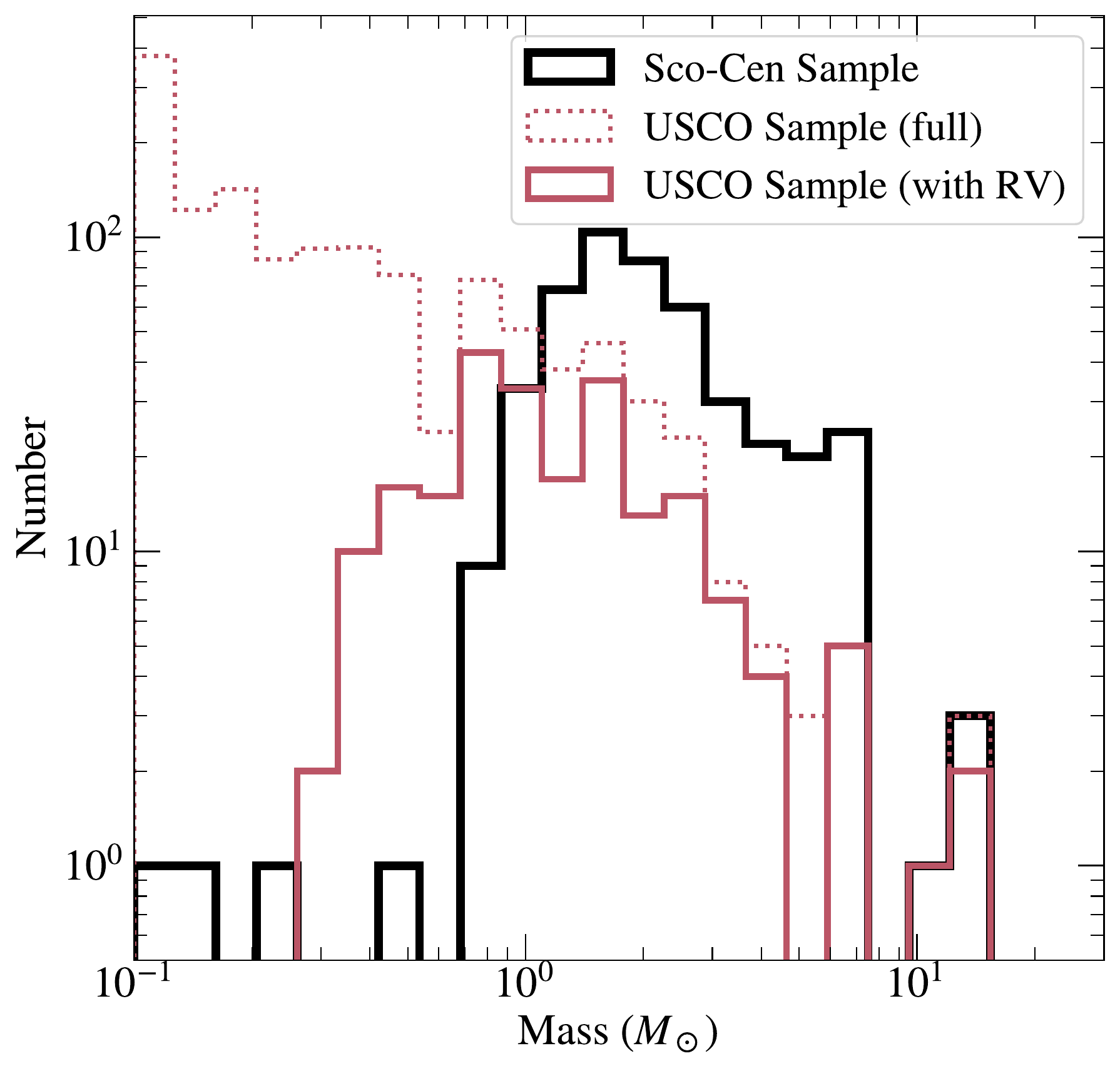}
   \caption{The mass distributions of the samples are plotted. The full samples are plotted with solid lines; the mass distributions of USCO stars with RV is plotted with the dotted line. The bin are drawn in the logarithmic space.}
    \label{fig:mass_distribution}
\end{figure}

This study used  the three-dimensional positions and motions of the known members of the nearby ($116-144\,\mathrm{pc}$, \citealt{1999AJ....117..354D}) Scorpius-Centaurus OB2 association to identify close stellar approaches. The positions and motions were derived from precise \textit{Gaia} Data Release 2 (DR2) astrometry \citep{gaia2018} and a combination of \textit{Gaia} and ground-based radial velocity (RV) measurements. The study focused on two samples of stars from this association: the first based on a census of stars in the region from the \textit{Hipparcos} satellite \citep{1997ESASP1200.....E}, and the second based on a revised census of the Upper Scorpius (US) subgroup using data from \textit{Gaia} DR2. 

\par The first sample, which we call the ``Sco-Cen'' sample, was a subset of a catalog of Sco-Cen members constructed based upon \cite{2018MNRAS.476..381W} and \cite{2012AJ....144....8S}, which contains 539 stars from all three subgroups of the association --- Upper Scorpius (USCO), Upper Centaurus-Lupus (UCL), and Lower Centaurus-Crux (LCC).
The subset we used in this study contains 462 catalog stars that have a radial velocity measurement from either \textit{Gaia} or ground-based instruments. When a target had a RV measurement from both \textit{Gaia} and other facilities, we opted to use the measurement with the lowest uncertainty. A full listing of the Sco-Cen sample is given in Table \ref{table:462scocen}.

\par The second group, which we call the ``USCO'' sample, is derived from 1682 stars in the \textit{Gaia} DR2 catalog that are identified as USCO members in Table 4 of \cite{luhman20}. We excluded 120 stars missing parallax measurements and 2 stars with negative parallax measurement and obtained a reduced sample of 1560 stars. This sample included 56 stars which are in the Sco-Cen sample described above. Among the stars in the USCO sample, 219 have RV from either the \textit{Gaia} catalog or ground-based measurements, and we adopt the same procedure for targets with RV measurements from both \textit{Gaia} and other facilities as with the other sample. A full listing of the USCO sample is given in Table \ref{table:usco_sample}. In total the two samples consist of 625 unique targets.

\subsection{Mass Estimation}\label{sec:mass_estimation}
\par To assess the relative significance of stellar flybys identified in this study, we required an estimate of the mass of each star within both samples. We generated color-magnitude diagrams (CMDs) of the two samples using optical photometry from \textit{Gaia}. Plotted in Fig. \ref{fig:CMD}, these clearly demonstrate that the Sco-Cen sample is restricted to more luminous stars ($M_G\lesssim 7$) due to the sample being primarily derived from \textit{Hipparcos} astrometry, whereas the USCO sample includes stars that are as faint as $M_G\sim15$. To assess the completeness of the two samples, we derived the stellar masses of the two samples based on the stars’ spectral types. We first interpolated a relation between spectral type and effective temperature based on Table 5 of \cite{2013ApJS..208....9P}. We then modelled the isochrones for the three subgroups with the MESA Isochrones and Stellar Tracks (MIST) \citep{2011ApJS..192....3P, 2013ApJS..208....4P, 2016ApJS..222....8D, 2016ApJ...823..102C}, assuming the ages of USCO, UCL, and LCC to be 10 Myr, 16 Myr, and 15 Myr, respectively \citep{2016MNRAS.461..794P}, and the metallicities of the subgroups to be solar for simplicity \citep{2009A&A...501..965V}. We obtained a relation between effective temperature and stellar mass, and thus between spectral type to stellar mass for each subgroup individually. For stars whose derived effective temperature is not covered by the isochrones, we used a luminosity-mass relation from the MIST isochrones and estimated their stellar masses using their G-band magnitudes in \textit{Gaia} DR2. We did not apply any extinction correction to the optical magnitudes as we do not require precise masses, although the extinction towards many of the sources in both samples is expected to be small (e.g. \citealp{2016MNRAS.461..794P}).

We generated the mass distribution of the two samples using the derived masses, which is shown in Fig. \ref{fig:mass_distribution}. For stars whose derived effective temperatures are so low that they exceed the isochrones' coverage, we assumed the mass to be $0.1\,M_\odot$, which causes a slight over-abundance of stars at the low-mass end of the mass distribution for the USCO sample. Based on the mass distributions, we found that the USCO sample is indeed bottom-heavy and matches the IMF quite well, while the Sco-Cen sample is deficient of lower-mass stars. The Sco-Cen sample is complete to $\sim1M_\odot$, and the USCO sample indeed includes more low-mass stars and is complete down to $\sim0.1M_\odot$. However, although the USCO sample has more stars with RV measurements at the low-mass end than the Sco-Cen sample does, the RV measurement is still incomplete for the star with $M\lesssim0.3\,\mathrm{M_\odot}$.

\par Ideally, we would have a more complete sample for the LCC and UCL subgroups as well in order to trace the dynamics of the subgroups more comprehensively. However, since a thorough census of these two sub-groups had not yet been performed using \textit{Gaia} DR2 astrometry, we decided not to mix the two samples and treat them separately to avoid bias against the two less-sampled subgroups.

\section{Methods}\label{sec:flyby_method}
\subsection{Flyby Identification}\label{sec:flyby_identification}
The astrometry and radial velocities are used to trace the stellar motion of the stars in both samples from 5 million years in the past to 2 million years in the future. We assumed that the stars are gravitationally unbound with each other and the time interval integrated over is short comparing to the galactic orbital periods of the stars so that the effect of the galactic potential can be neglected. Therefore, the trajectories of the stars can be approximated as linear. In their analysis of stellar flybys of HD 106906, \cite{derosa_kalas} showed that there is not a significant deviation in the result between the linear approximation and an N-body simulation. We assumed that this is true for all of the stars within both samples. Based on the linear approximation, we computed the time of closes approach $t_\mathrm{ca}$ between any pair of stars in the sample as
\begin{equation}\label{t_ca}
    t_{\text{ca}} = -\frac{(\vec{r_2}-\vec{r_1})\cdot(\vec{v_2}-\vec{v_1})}{\left|\vec{v_2}-\vec{v_1}\right|^2}\,\,,
\end{equation}
where $\vec{r_1}$, $\vec{r_2}$ and $\vec{v_1}$, $\vec{v_2}$ represent the pairs of positions and velocities of the two stars. Subsequently, we are able to compute the distance of closest approach $d_{\rm ca}$ of each pair of stars as 
\begin{equation}\label{d_ca}
    d_{\text{ca}} = \left|(\vec{r_2}-\vec{r_1})+(\vec{v_2}-\vec{v_1})t_{\text{ca}}\right|\,\,.
\end{equation}

\par Uncertainties of the \textit{Gaia} astrometry and the radial velocity (RV) measurement are propagated with a Monte Carlo (MC) fashion. For each of the $10^6$ trials, we drew the position, parallax, and proper motion of both stars based on a multivariate normal distribution using the correlation coefficients in the \textit{Gaia} DR2 catalog and drew the radial velocities from a separate Gaussian distribution. Using Equation \ref{t_ca} and Equation \ref{d_ca}, $t_\text{ca}$ and $d_\text{ca}$ were computed for all $10^6$ trials. We used the median values of $t_\text{ca}$ and $d_\text{ca}$ as the adopted values for these parameters, and the 16th- and 84th-percentiles were used as the lower and upper bounds, respectively. 
\par We used the star's Hill radius $R_H$ under the galactic potential as a mass-weighted distance to classify the flybys. The Hill radius encodes the radial boundary of a volume where a star has a greater gravitational influence on an object than the galactic potential. The Hill radii were approximated using 
\begin{equation}\label{eq:hill}
    R_H \approx a\left(\frac{m_\star}{3M_\text{enc}(a)}\right)^{\frac{1}{3}}\,\,,
\end{equation}
where $a$ is the distance from the galactic center to the star, $m_\star$ is the derived stellar mass, and $M_\text{enc}(a)$ is the mass enclosed within the star's galactic orbit assuming a flat rotation curve for our Galaxy. If $d_\text{ca}< 0.5 R_H$ for {\it either} star, it is categorized as a ``close encounter."  If $0.5R_H\leqslant d_\text{ca}\leqslant R_H$ (for either star), then the event is called an ``encounter." For more generic references to unbound interactions we use the term ``flyby."  We do not use these terms to refer to the periastron passages of bound objects even if the orbits are highly eccentric and with large semi-major axes. Tables 3--5 have columns $d_{ca}/R_{H,T}$ and $d_{ca}/R_{H,F}$ where the former lists how the closest approach distance compares to the Hill radius of the target (T) and the latter with respect to the Hill radius of the flyby star (F).

\subsection{Flyby Geometry}\label{sec:flyby_geometry}
For each flyby event that involve a resolved debris disk, we also derived the geometry of the closest approaches.
Assuming the disks are azimuthally symmetrical centered around their host stars, we used four reference points around the disk edge to carry out rotations that eventually leads to orienting the disks in galactocentric coordinates, the coordinate system we used for calculating the flyby timing and distances. These reference points are defined as 
\begin{equation}\label{four_points}
    \vec{p_1} = \begin{bmatrix} 0 \\ \frac{r}{d} \\ 0 \end{bmatrix}, \vec{p_2} = \begin{bmatrix} 0 \\ -\frac{r}{d} \\ 0 \end{bmatrix}, \vec{p_3} = \begin{bmatrix} 0\\ 0 \\ \frac{r}{d} \end{bmatrix}, \vec{p_4} = \begin{bmatrix} 0\\ 0 \\ -\frac{r}{d}\end{bmatrix},
\end{equation}
where $d$ is the distance to the host star and $r$ is a nominal radius of the disk measured from the resolved images. The origin (0, 0, 0) is defined to be the position of the host star, and the three axes are along the RA, DEC, and line-of-sight directions, respectively. We carried out the first rotation based on the position angle (PA) of the disk via the rotation matrix 
\begin{equation}\label{pa_rotation}
    \mathbf{R_{\boldsymbol{\theta}}} = 
    \begin{bmatrix}
        \cos(\text{PA}) & \sin(\text{PA}) & 0 \\ 
        -\sin(\text{PA}) & \cos(\text{PA}) & 0 \\
        0 & 0 & 1
    \end{bmatrix}.
\end{equation}
To rotate the reference points according to the inclination of the disk, we first defined the normal vector about which this second rotation was carried out as 
\begin{equation}\label{2nd_rot_axis}
    \mathbf{\hat{n}} = \mathbf{R_{\boldsymbol{\theta}}}\left(\frac{\vec{p_2}-\vec{p_1}}{|\vec{p_2}-\vec{p_1}|}\right).
\end{equation}
Thus, we constructed our inclination-rotation matrix as
\begin{equation}\label{inc_rotation}
    \vec{R_{\boldsymbol{i}}} =
    \begin{bmatrix}
        (1-s)+n_1^2s & n_1n_2s-n_2c & n_2n_3s+n_2c \\ 
        n_1n_2s+n_3c & (1-s)+n_2^2s & n_2n_3s-n_1c \\
        n_1n_3s-n_2c & n_2n_3s+n_1c & (1-s)+n_3^2s
    \end{bmatrix}\,\,,
\end{equation}
where $s=1-\sin{i}$, $c=\cos{i}$ and $n_1$, $n_2$, and $n_3$ are the three components of $\hat{\vec{n}}$, respectively. After correcting the projection on the sky plane, we add the offset of the stars' positions in Cartesian ICRS coordinates to all four points to transform from the star frame to the ICRS coordinate frame. We then transformed these four points from ICRS to galactocentric coordinates using the procedures defined in {\tt\string astropy.coordinates} . In the galactocentric coordinates, we define the normal vector of the debris disk as
\begin{equation}\label{disk_normal_vector}
    \hat{\vec{N}} = \frac{(\vec{p_1'}-\vec{p_2'})\times(\vec{p_3'}-\vec{p_4'})}{|(\vec{p_1'}-\vec{p_2'})\times(\vec{p_3'}-\vec{p_4'})|}\,\,, 
\end{equation}
where the primed vectors are the galactocentric coordinates of the reference points. We then calculated the closest approach angle $\theta_{\text{ca}}$, the angle between the disk plane and the flyby star's velocity vector ($\vec{v_{\text{rel}}}$) in the rest frame of the target star, as
\begin{equation}\label{flyby_angle}
    \theta_{\text{ca}} = \arcsin\left(\frac{\hat{\vec{N}}\cdot\vec{v_{\text{rel}}}}{|\vec{v_{\text{rel}}}|}\right)\,.
\end{equation}
The uncertainties on all properties were propagated in a Monte Carlo fashion. The angle of closest approach defined previously is calculated for each of the $10^6$ draws, where we drew disk position angles and inclinations from Gaussian distributions using the literature measurement and uncertainty. We validated our implementation of this algorithm by comparing to the results for HD 106906 presented in \cite{derosa_kalas}.

\section{Results}\label{results}
\subsection{Flyby Statistics}\label{sec:flyby_statistics}

\begin{figure}
\includegraphics[width=\columnwidth]{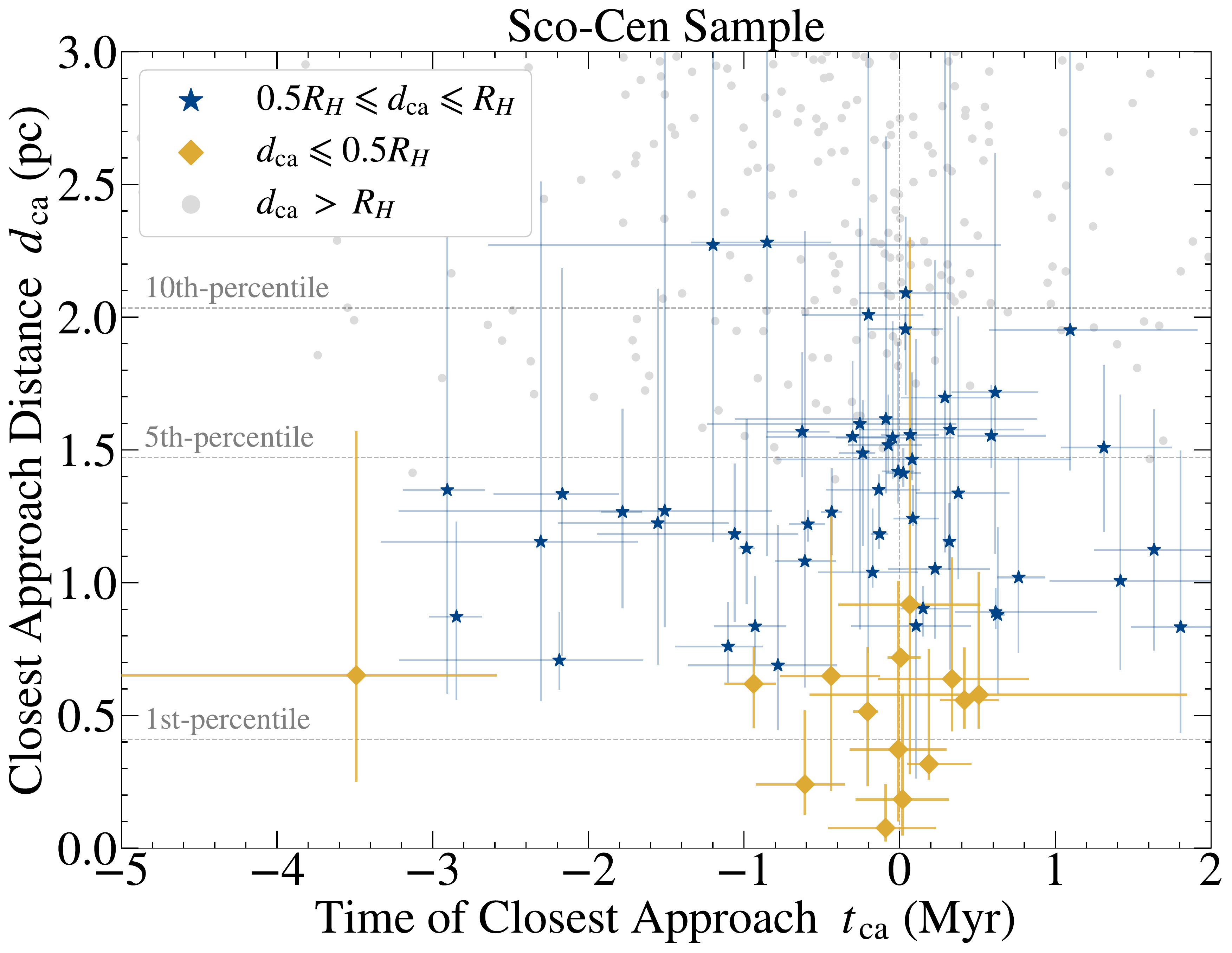}
\caption{Closest approach times and distances for the encounters and close encounters identified in the Sco-Cen sample. Orange diamonds mark the close encounters ($d_\text{ca}\leqslant0.5R_\text{H}$); blue stars mark encounters ($0.5R_\text{H} \leqslant d_\text{ca}\leqslant R_\text{H}$); gray dots mark the pairs whose close approach distance is in neither star's Hill radius. The dotted horizontal lines mark the 1st-, 5th- and, 10th-percentiles of the distribution of the separation of stars within the sample. The median separation of stars in the Sco-Cen sample is 5.60 pc.}
\label{fig:dtscocen}
\end{figure}

\begin{figure}
    \includegraphics[width=\columnwidth]{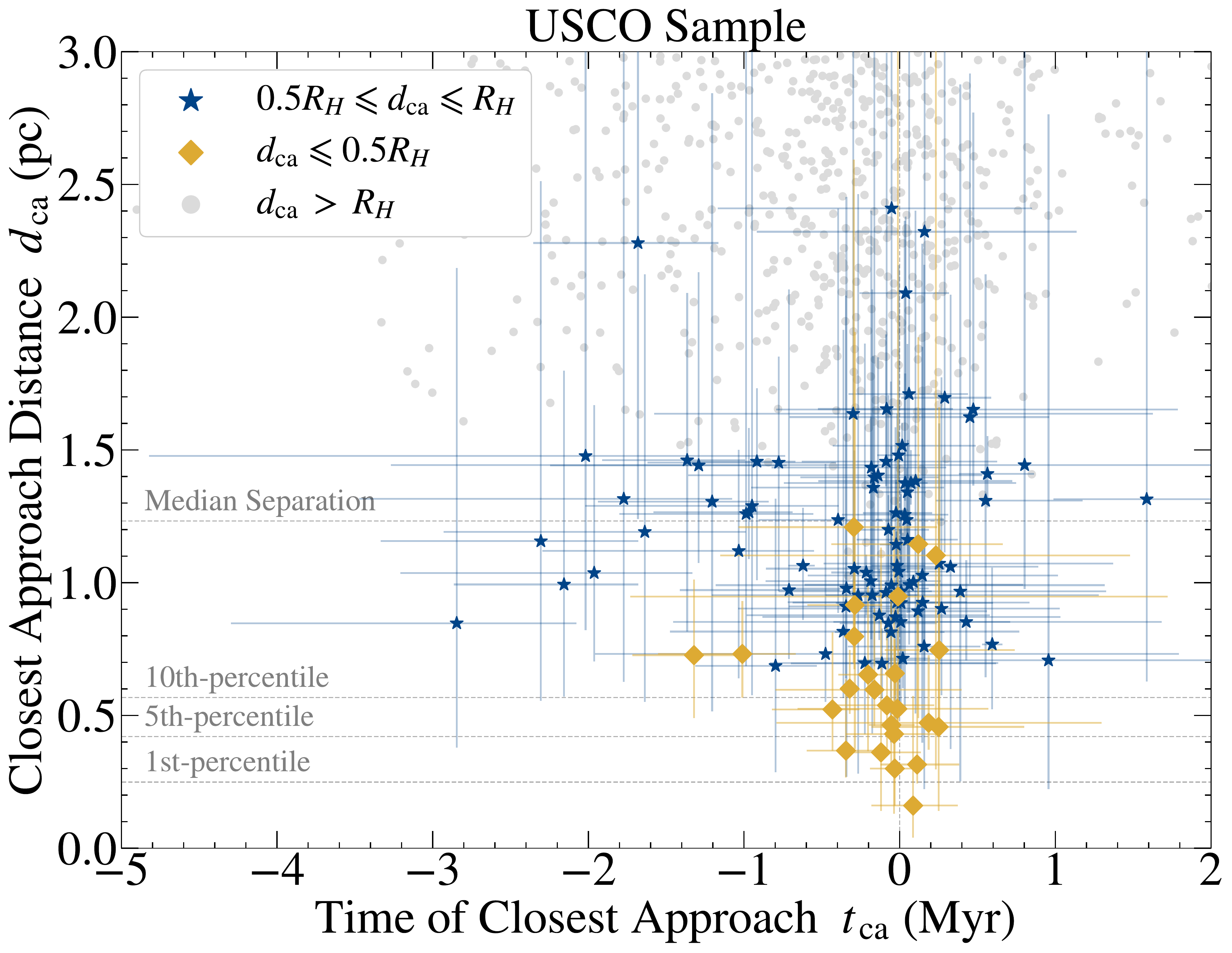}
    \caption{Closest approach times and distances for all the encounters and close encounters identified in the USCO sample. Legends are the same as that in Fig.\ref{fig:dtscocen}. The median separation of stars in this sample is 1.23 pc.}
    \label{fig:luhman_dtplot}
\end{figure}

In the Sco-Cen sample, we identified seven past $close$ encounters and seven future close encounters, each involving 13 stars, along with 37 past encounters and 28 future encounters involving 63 and 45 stars, respectively. There are 98 individual stars involved in all these events (including both encounters and close encounters). Similarly, we carried out the same calculation on 219 stars in the USCO sample where we identified 18 past close encounters involving 33 stars, 7 future close encounters involving 13 stars, 60 past encounters involving 81 stars, and 39 future encounters involving 64 stars. In total, 116 individual stars are involved in all these events. The time and distance of closest approach events of both samples are presented in Figures \ref{fig:dtscocen} and \ref{fig:luhman_dtplot}. As the time approaches the present, more encounters and close encounters are identified. We attribute this phenomenon to the increasing uncertainties of $d_\text{ca}$ and $t_\text{ca}$ as we trace further back to the past and into the future due to the non-zero uncertainties of the astrometry and radial velocities, an effect also seen in studies of stellar encounters with our own solar system (e.g., \citealp{bailer-jones18a}). The full list of flyby events is presented in Table \ref{table:all_scocen_encounters}. 

\begin{figure*}
    \centering
    \includegraphics[width=\textwidth]{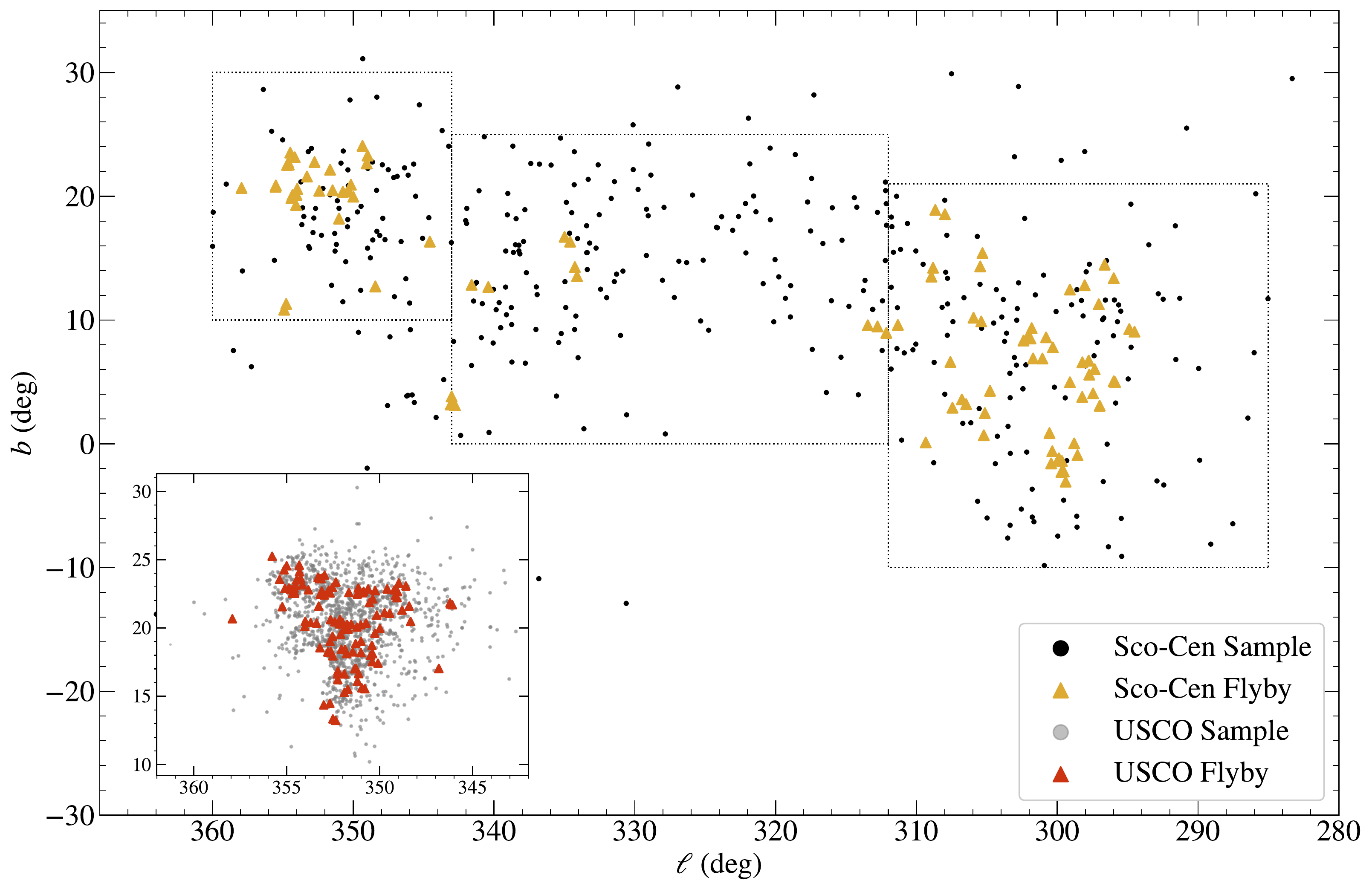}
    \caption{The spatial distribution of the two samples as well as the stars with identified flyby events is plotted in the Galactic coordinates. The dashed boxes mark the boundaries of USCO, UCL, and LCC identified by \cite{1999AJ....117..354D}. The USCO sample is plotted within the inset.}
    \label{fig:scocen_in_sky}
\end{figure*}
\par The spatial distribution of flybys in both samples is presented in Fig. \ref{fig:scocen_in_sky}. We observe that stars involved in flybys are distributed evenly in the sky, except for a deficiency of events between $\ell=314^\circ$ and $\ell=334^\circ$. This gap in the spatial distribution of flybys is coincident with a lower density of stars seen on the western side of the UCL subgroup. The lower density within this region is also seen in the spatial distribution of fainter pre-main sequence candidate members identified from an analysis of \textit{Gaia} DR2 data (e.g., \citealp{2019A&A...623A.112D}). The lower occurrence of identified flybys within this region can simply be ascribed to a lower three-dimensional density of stars within this part of the association.

\par The algorithm we used to identify flybys would naturally include some wide binaries due to the fact that they share similar positions and velocities. According to the Washington Double Star Catalog (WDS; \citealp{2020yCat....102026M}), in the Sco-Cen sample, 22 out of the 98 Sco-Cen stars involved in flybys have binary companions; 16 out of the 116 USCO stars involved in flybys have binary companions as well; however, none of the identified flyby pairs is listed as physically associated. Nonetheless, HD 143844 and HD 143215, which are identified to have a flyby event, are identified as wide binaries in \cite{2020ApJS..247...66H}. Four other flyby events that we identified (HD 144587 / HD 144175, HD 146897 / HD 147083, HD 137499 / HD 137432, and HD 121835 / HD 121190) are also catalogued to be wide binary systems by \cite{2019AJ....157...78J}. However, it is very difficult to determine whether a wide pair is gravitationally bound given the very low relative orbital velocities and the current radial velocity uncertainties for many of these stars. It is also worth noting that when tracing the trajectories, the radial velocity measurements of some stars that are in short-period binary systems may not be representative of the systematic velocity, which would cause an incorrect determination of the three-dimensional trajectories of these stars. Where possible we used systemic velocities for spectroscopic binaries derived from orbit fits presented in the literature (e.g., \citealp{1987ApJS...64..487L}).


\subsection{Field Star Flybys}\label{sec:field_stars}
We also ran our closest approach algorithm on a dataset combining all stars in the \textit{Gaia} DR2 that were not a part of either of the Sco-Cen samples that have RV measurements from either \textit{Gaia} or other ground-based observations reported in SIMBAD. Among this dataset, we flagged the stars that are identified by \cite{2019A&A...623A.112D} as candidate Sco-Cen members based on their {\it Gaia} DR2 astrometry. Those that are not flagged are hereafter referred to as the field sample or field stars. This field sample consists of approximately 7.5 million stars, with greater completeness than the Sco-Cen sample to low-mass stars (0.5--1\,$M_{\odot}$) owing to the magnitude and effective temperature range of stars with {\it Gaia} DR2 radial velocities.

\begin{figure}
    \includegraphics[width=\columnwidth]{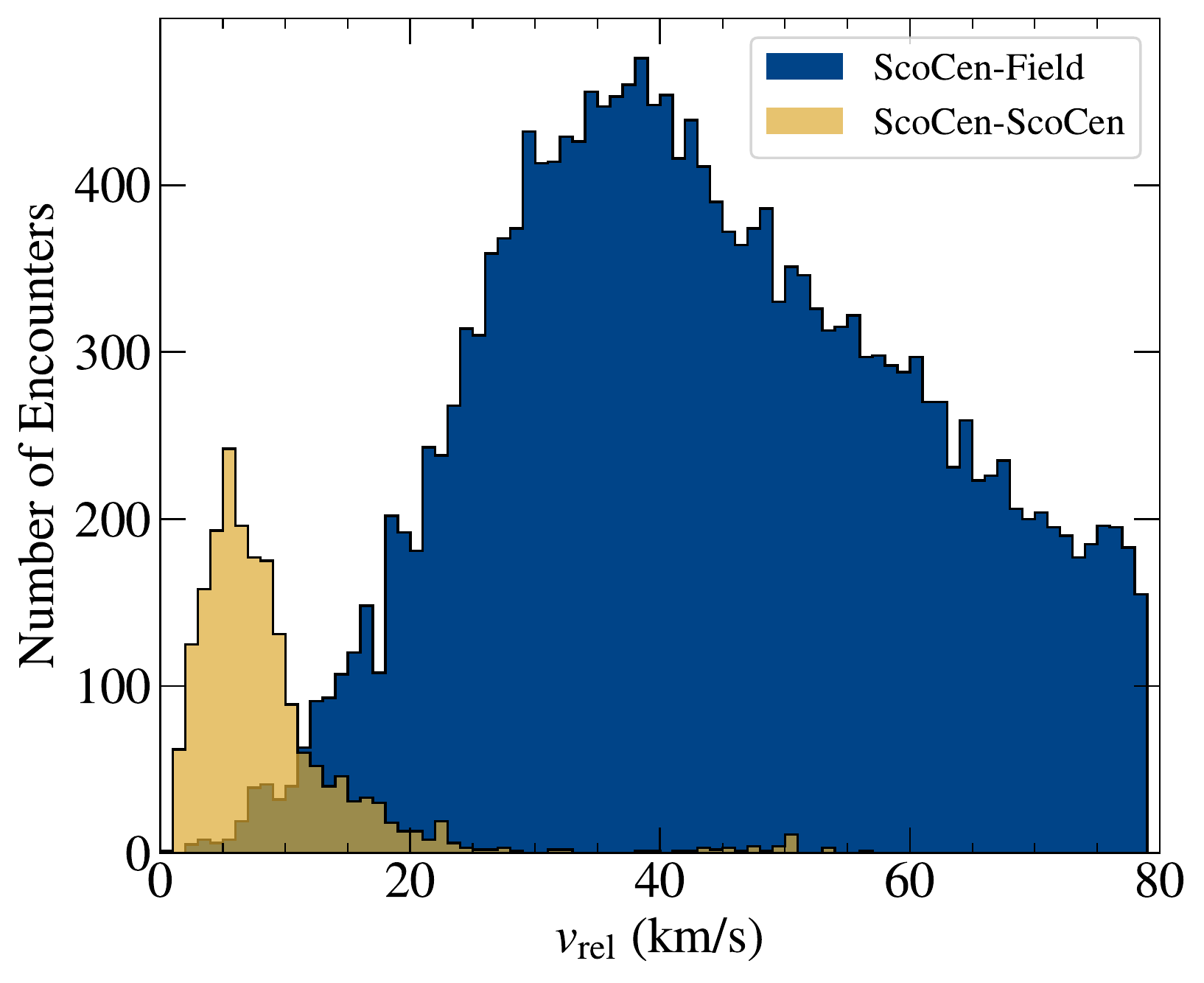}
    \caption{The comparison of relative velocity distributions between ScoCen-Field encounters (blue) and ScoCen-ScoCen encounters (orange) with $d_\mathrm{ca}<3\,\mathrm{pc}$.}
    \label{fig:vrel_hist}
\end{figure}

We found that flybys involving field stars with $d_{\rm ca}<3$\,pc are significantly more frequent than those only involving members of the Sco-Cen sample, in large part due to the difference in completeness between the two samples. However, the two types of flyby (ScoCen-Field vs. ScoCen-ScoCen) have different relative velocity distributions as shown in Fig.\ref{fig:vrel_hist}. The median relative velocity for encounters between Sco-Cen members and field stars is $\sim45\,\mathrm{km/s}$, whereas that for encounters between two Sco-Cen members is $\sim7\,\mathrm{km/s}$.

\begin{figure}
    \centering
    \includegraphics[width=\columnwidth]{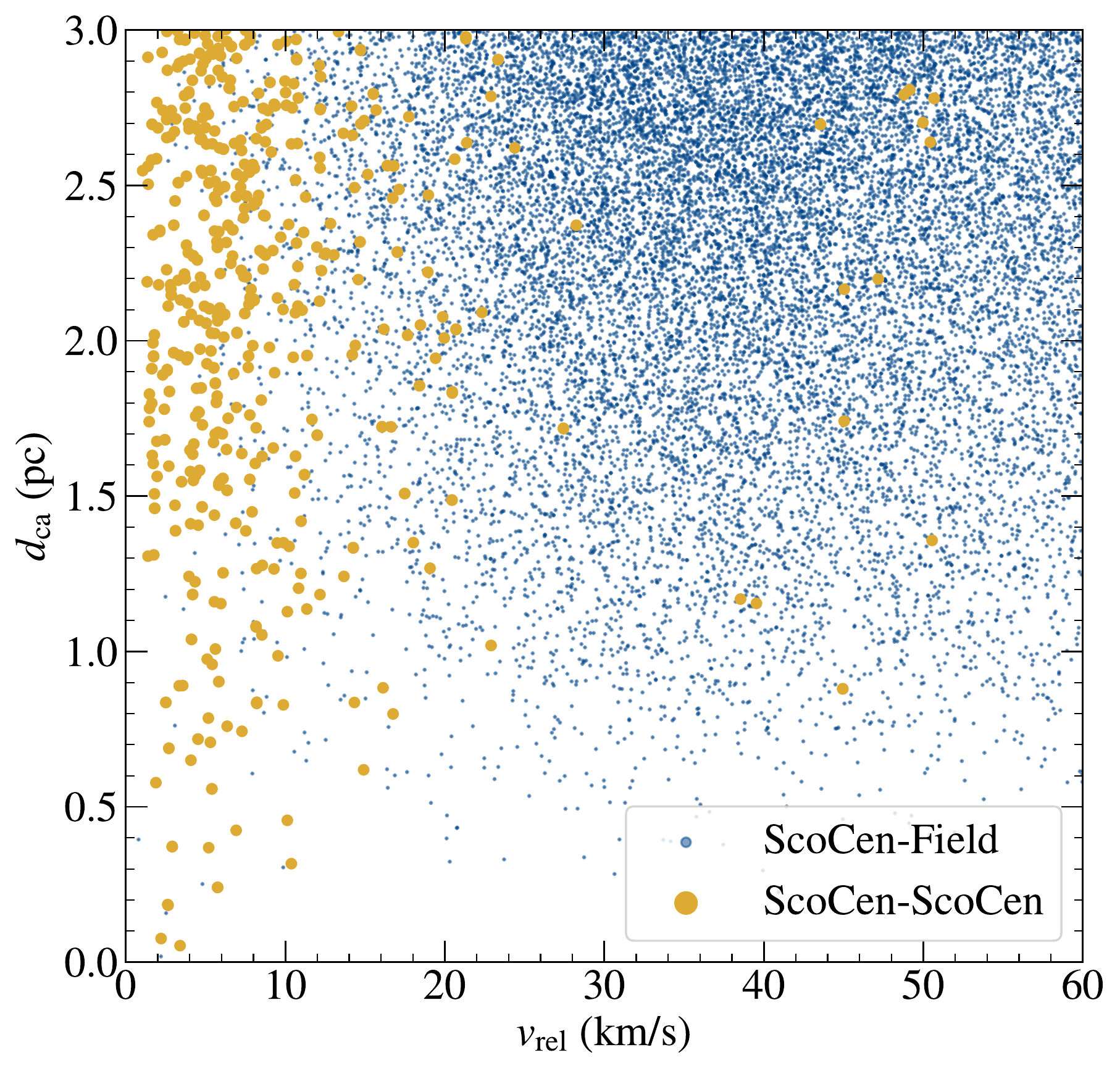}
    \caption{The closest approach distance is plotted against relative velocity for ScoCen-field encounters (blue) and ScoCen-ScoCen encounters (orange).}
    \label{fig:dca_vrel_field}
\end{figure}

If we use the change in orbital speed as a measure for dynamical influence (e.g., \citealp{2001ApJ...553..410K}), we see this difference in relative velocity causes a factor of $\sim6$ difference between the two types of encounters. Furthermore, it can be observed in Fig.\ref{fig:dca_vrel_field} that the events with $d\lesssim1\,\mathrm{pc}$ and $v_\mathrm{rel}\lesssim10\,\mathrm{km/s}$ are dominated by encounters between members of the association. From this we infer that it is the encounters between members of the association that are more dynamically important than with field stars. Nevertheless, we report the details of encounters with field stars, and with candidates from \citet{2019A&A...623A.112D}, in Table \ref{table:candidate_field}.


\subsection{Correlation with infrared excess} \label{sec:ir_excess}
Various effects are possible when close stellar flybys interact with a star surrounded by a disk of planetesimals and/or planet-mass objects.  The flyby can strip away outer material, truncating a circumstellar disk, but also increasing the volume density of surviving material and dynamically heating it \citep{pfalzner05a, lestrade11a}.  This would increase the rate of dust-producing collisions, manifesting as an infrared excess until the entire system dynamically relaxes and dust grains are depleted by collisional destruction, sublimation, or removal by radiation forces. The 5 Myr-old HD 141569 debris disk may be undergoing this process at the current epoch due to the flyby of one or more M stars \citep{weinberger00a,augereau04a,ardila05a}.
To test for a possible correlation between flybys and IR excesses, we crosschecked both of our samples with the list of IR excess stars compiled in \cite{cotten2016} and determined the fraction of the sample stars with excess. In the Sco-Cen sample, we identified 126 stars with IR excess with the dust temperature between 50K and 575K; 33 out of 67 events involve 24 IR-excess stars. In the USCO sample, there are 42 IR-excess stars with the dust temperature ranging from 75 K to 450 K; 21 out of 117 events involve 9 IR-excess stars. 

\par
\begin{figure}
    \centerline{\includegraphics[width=250pt]{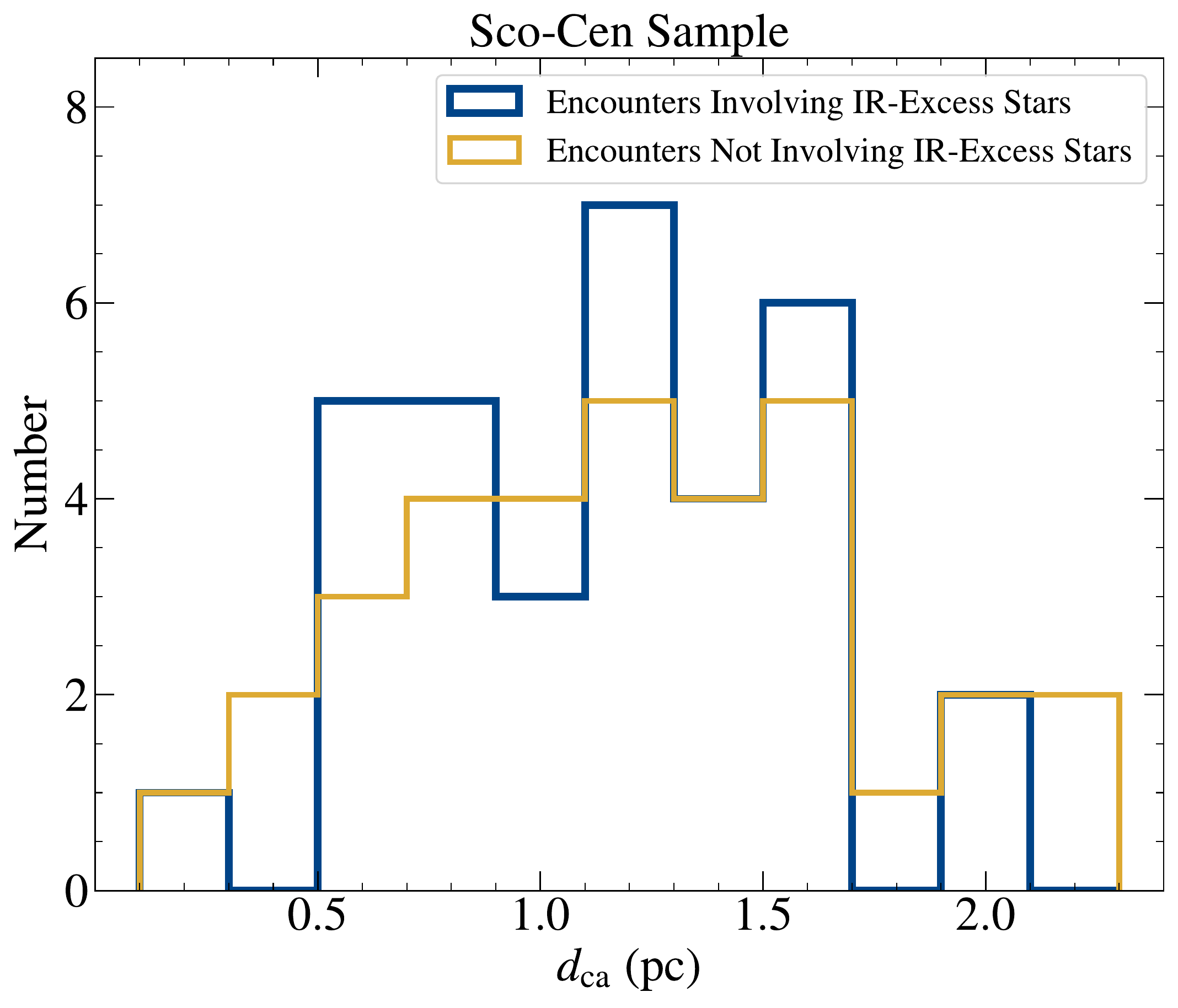}}
    \caption{The comparison of the close approach distances of the events involving and not involving IR-excess stars for the Sco-Cen sample.}
    \label{fig:ir_correlation}
\end{figure}

 As shown in Figure \ref{fig:ir_correlation}, the distributions of the close approach distance for stars with and without IR excess overlap for the Sco-Cen sample. However, since the \cite{cotten2016} study does not have a uniform sensitivity for the IR excess detection for all the stars and the sample itself is less complete for the low mass stars, we would conservatively conclude that our data do not provide evidence for a correlation between the presence of IR excess and stellar flybys. For the same issue of sensitivity, only 42 out of 1560 stars in the USCO sample are included in \citet{cotten2016}. Most stars in the USCO sample were too faint to be catalogued and have only been recently identified with data from the \textit{Gaia} satellite, which makes the sample bias more significant. Consequently, we did not include USCO sample in this analysis. 

\subsection{Flybys of stars with disks detected in scattered light or with ALMA} \label{sec:interaction}

To search for morphological evidence of dynamical interactions between stellar flybys and circumstellar material (e.g., radial asymmetries, spiral structures, and vertical warps), we surveyed the literature for resolved detections of debris disks around the stars identified as having flyby events with other association members in the past. Table \ref{table:geometry} and Fig. \ref{fig:each_geometry} present the flyby parameters, including the current 3D separations between stars ($d_0$), the relative velocities of each flyby ($v_\text{rel}$), and the relative angles between the disk midplanes and flyby trajectories ($\theta_\text{ca}$).

Out of all the flyby stars from both samples, five have debris disks resolved in \emph{scattered light} at visible and/or near-infrared wavelengths: HD 106906, HD 114082, HD 115600, HD 146897, and 2MASS J16042165-2130284 (RXJ1604.3-2130). At thermal wavelengths, the Atacama Large Millimeter/Submillimeter Array
(ALMA) has been used to resolve the disks surrounding 2MASS J16064794-1841437 and HD 145718 \citep{ansdell20a}.
In the USCO sample, 2MASS J16035767-2031055 hosts a circumstellar disk \citep{2012ApJ...758...31L} also resolved with ALMA at 0.88 mm continuum and the CO $J=3-2$ line at 345.8 GHz  \citep{2016ApJ...827..142B}.
Circumstellar dust disks were weakly detected (SNR$\sim$3--4) but unresolved at 1.24 mm by \cite{LS2016} using ALMA with $\sim$1'' resolution  around HD 113556 and HD 142315.
We include these two disks in our discussion to anticipate the availability of higher resolution continuum maps in the near future. 
We discuss the details of all 10 cases below.

\begin{deluxetable*}{cccccccc}
\tablecaption{Flyby events involving resolved or detected debris disks\label{table:geometry}}
\tablehead{\colhead{Target star} & \colhead{Flyby star} & \colhead{$d_0$} & \colhead{$d_{\text{ca}}$} & \colhead{$t_{\text{ca}}$} & \colhead{$\theta_\text{ca}$} & \colhead{$v_\text{rel}$} & \colhead{Ref}\\ 
\colhead{} & \colhead{} & \colhead{(pc)} & \colhead{(pc)} & \colhead{(Myr)} & \colhead{(deg)} & \colhead{(km/s)} & \colhead{}}
\startdata
HD 106906& HD 106444 & $11.55_{-0.68}^{+0.69}$ & $0.65_{-0.40}^{+0.92}$ & $-3.49_{-1.75}^{+0.90}$ & $5.90_{-1.67}^{+1.77}$ & $3.22_{-1.09}^{+1.10}$&1\\
{} & HIP 59721 & $11.89_{-0.65}^{+0.64}$ & $0.71_{-0.11}^{+0.18}$ & $-2.19_{-1.03}^{+0.54}$ & $4.64_{-1.16}^{+1.00}$ & $5.30_{-1.69}^{+1.70}$&\\\hline
HD 113556 & TYC 8674-2317-1 & $7.03_{-0.46}^{+0.47}$ & $1.22_{-0.53}^{+0.88}$ & $-1.55_{-0.64}^{+0.46}$ & \nodata$\dagger$ & $4.34_{-1.45}^{+1.65}$&2\\\hline
HD 114082 & HD 116402 & $26.09_{-0.86}^{+0.87}$ & $0.87_{-0.31}^{+0.36}$ & $-2.85_{-0.17}^{+0.16}$ & $-13.91_{-3.98}^{+3.96}$ & $8.95_{-0.40}^{+0.40}$&1\\\hline
HD 115600 & TYC 8674-2317-1 & $2.37_{-0.54}^{+0.56}$ & $0.89_{-0.06}^{+0.09}$ & $0.62_{-0.26}^{+0.65}$ & $10.52_{-1.69}^{+1.79}$ & $3.35_{-1.79}^{+1.81}$&1\\\hline
HD 142315 & HD 142705 & $2.21_{-0.53}^{+1.27}$ & $1.60_{-0.77}^{+0.77}$ & $-0.26_{-0.98}^{+0.62}$ & \nodata$\dagger$ & $2.70_{-1.34}^{+2.52}$&2\\
{} & UCAC4 336-078484 & $1.93_{-0.47}^{+1.10}$ & $1.36_{-0.45}^{+0.50}$ & $-0.17_{-0.79}^{+0.46}$ & \nodata$\dagger$ & $3.40_{-2.02}^{+3.47}$&\\
{} & HD 142097 & $4.91_{-1.26}^{+1.35}$ & $1.55_{-0.12}^{+0.19}$ & $0.59_{-0.22}^{+0.35}$ & \nodata$\dagger$ & $7.77_{-2.59}^{+2.67}$&\\\hline
2M J16035767-2031055 & 2M J16053815-2039469 & $5.24_{-1.23}^{+1.24}$ & $0.52_{-0.16}^{+0.29}$ & $-0.43_{-0.39}^{+0.17}$ & $-11.52_{-27.92}^{+29.51}$ & $11.37_{-5.66}^{+5.81}$&3\\
{} & RX J1603.9-2031B & $9.05_{-4.91}^{+5.29}$ & $0.71_{-0.48}^{+2.06}$ & $0.96_{-1.23}^{+1.80}$ & $-11.65_{-29.84}^{+33.98}$ & $6.29_{-4.19}^{+5.68}$&\\\hline
2M J16042165-2130284 & EPIC 204548337 & $1.98_{-0.73}^{+1.35}$ & $1.06_{-0.69}^{+1.03}$ & $0.33_{-0.59}^{+0.56}$ & $-27.59_{-47.42}^{+86.39}$ & $2.86_{-1.18}^{+2.44}$&4\\\hline
2M J16064794-1841437 & HD 144925 & $18.19_{-1.72}^{+1.72}$ & $0.73_{-0.24}^{+0.28}$ & $-1.32_{-0.40}^{+0.26}$ & $-30.90_{-0.83}^{+1.21}$ & $13.46_{-2.95}^{+2.96}$&5\\
\hline
HD 145718 & HD 145467 & $14.04_{-2.05}^{+2.11}$ & $1.16_{-0.60}^{+1.36}$ & $-2.31_{-1.03}^{+0.63}$ & $-20.56_{-1.84}^{+1.87}$ & $5.96_{-1.82}^{+1.86}$ &5\\
{} & 2M J16120920-2247504 & $11.14_{-2.17}^{+2.21}$ & $1.04_{-0.33}^{+0.63}$ & $-1.96_{-1.25}^{+0.63}$ & $-18.17_{-1.81}^{+2.11}$ & $5.50_{-2.07}^{+2.08}$ &\\
{} & BD-21 4301 & $18.21_{-2.58}^{+2.57}$ & $1.30_{-0.79}^{+1.54}$ & $-1.20_{-0.73}^{+0.36}$ & $-19.15_{-1.40}^{+1.51}$ & $14.66_{-5.57}^{+5.61}$ &\\
{} & Gaia DR2 6242198746659039872 & $8.66_{-2.05}^{+2.10}$ & $1.12_{-0.48}^{+0.38}$ & $-1.03_{-1.26}^{+0.48}$ & $-21.61_{-2.68}^{+2.01}$ & $7.32_{-4.41}^{+4.78}$ &\\
{} & 2M J16125889-2245202 & $2.86_{-1.69}^{+2.20}$ & $0.93_{-0.18}^{+0.96}$ & $0.15_{-0.83}^{+0.69}$ & $8.91_{-21.88}^{+31.88}$ & $4.70_{-2.99}^{+4.93}$&\\
\hline
HD 146897 & HD 146416 & $2.00_{-0.57}^{+1.16}$ & $0.65_{-0.43}^{+0.61}$ & $-0.44_{-0.33}^{+0.31}$ & $4.23_{-4.34}^{+4.37}$ & $4.08_{-0.32}^{+0.33}$&1\\
{} & HD 147083 & $1.82_{-1.03}^{+1.59}$ & $0.58_{-0.13}^{+0.46}$ & $0.51_{-1.09}^{+1.34}$ & $5.19_{-12.69}^{+10.26}$ & $1.88_{-1.21}^{+1.76}$&\\
{} & HD 146743 & $6.62_{-1.27}^{+1.27}$ & $1.12_{-0.38}^{+0.53}$ & $1.63_{-0.39}^{+0.49}$ & $-2.26_{-4.20}^{+4.36}$ & $3.91_{-0.75}^{+0.75}$&
\enddata
\tablecomments{$\dagger$ The target star disks have not been spatially resolved.}
\tablerefs{(1) \citealp{esposito_2020}; (2) \citealp{LS2016}; (3) \citealp{2016ApJ...827..142B, 2017ApJ...851...85B}; (4) \citealp{mayama12a}; (5) \citealp{ansdell20a}}
\end{deluxetable*}

\subsubsection{HD 106906}\label{dis:106906}
We confirm the \cite{derosa_kalas} result that HD 106906 (F5V) has experienced two flyby events: one close encounter with HD 106444 (F5V) at $d_{\text{ca}}= 0.65_{-0.40}^{+0.92}\,\text{pc}$ and $t_{\text{ca}}= -3.49_{-1.75}^{+0.90}\,\text{Myr}$ and an encounter with HIP 59721 (G9V) at $d_{\text{ca}}= 0.71_{-0.11}^{+0.18}\,\text{pc}$ and $t_{\text{ca}}= -2.19_{-1.03}^{+0.54}\,\text{Myr}$. The two candidate perturbers likely form a wide binary system \citep{mason01a}. The two events are only separated by $\sim1.5\,\text{Myr}$ and are nearly co-planar, the closest approach angles are $5.90_{-1.67}^{+1.77}$ deg and $4.64_{-1.16}^{+1.00}$ deg, respectively. 

An asymmetric inner disk is observed in $H$-band scattered light \citep{kalas_2015, lagrange16a}. Moreover, \cite{kalas_2015} detected a significant asymmetry of the outer disk ($>$200 au projected radius) via \textit{Hubble Space Telescope}/Advanced Camera for Surveys (\textit{HST}/ACS), with a truncated fan structure on the southeast side and an extended linear morphology to the northwest. The outer giant planet HD 106906 b is a likely source of the perturbation, with the flybys playing a dynamical role in raising its periastron distance away from the central region close to the binary \citep{rodet17a, nesvold17a, 2021AJ....161...22N}. 

\subsubsection{HD 113556}\label{dis:113556}
We found an encounter between HD 113556 (F2V) and TYC 8674-2317-1 (K4) with $d_{\text{ca}} = 1.22_{-0.53}^{+0.88}\,\text{pc}$ at $t_\mathrm{ca} = -1.55_{-0.64}^{+0.46}\,\mathrm{Myr}$. HD 113556 has Spitzer-detected 24 and 70 $\mu$m infrared excesses \citep{chen05a, chen11a} in addition to the ALMA detection at 1.24 mm \citep{LS2016}, which is unresolved. The flyby star, TYC 8674-2317-1, does not have reported thermal-IR excess from its \textit{WISE} photometry, although it has not been targeted with either \textit{Spitzer}, \textit{Herschel}, or ALMA, all of which are more sensitive to cold dust at wider separations. 

\subsubsection{HD 114082}\label{dis:114082}
HD 114082 (F3V) had one encounter $2.85_{-0.17}^{+0.16}$ Myr ago with $d_\text{ca}=0.87_{-0.31}^{+0.36}$ pc. The interaction has  $\theta_{\text{ca}}=-13.91^{+3.96}_{-3.98}\,\text{deg}$ and a high relative velocity ($v_\text{rel}=8.95\pm0.40\,\text{km/s}$). \citet{2016A&A...596L...4W} first resolved the HD 114082 disk in $H$-band scattered light as a ring with inner radius $\sim$28 au and $i\sim83\degr$. \citet{esposito_2020} confirmed the ring structure in both $H$-band total intensity and polarized intensity images.  They noted that dust scattered light was 1.5$-$1.8 times brighter on the west side of the disk compared to the east out to a radius of 1.2". Deep optical images obtained with HST/STIS have detected the disk ansae to $\sim$5.4'' (520 au) radius from the star but a significant asymmetry in brightness or structure between the two ansae is not evident (T. Esposito, priv. comm.). The flyby star, HD 116402 (G3V), does not have a measured infrared excess, and there are no published high-contrast imaging data searching for a debris disk around this star.

\subsubsection{HD 115600}\label{dis:115600}
We calculated that HD 115600 (F2IV/V) will encounter TYC 8674-2317-1 with $t_{\text{ca}}=0.62_{-0.26}^{+0.65}\,\text{Myr}$ and $d_\text{ca}=0.89_{-0.06}^{+0.09}\,\text{pc}$.  Though $t_{\text{ca}}$ is in the future, a closest approach at the present epoch is within the 3-$\sigma$ uncertainty.  The closest approach angle is $\theta_{\text{ca}}=10.52_{-1.69}^{+1.79}\,\text{deg}$. Previous GPI \citep{2015ApJ...807L...7C} and SPHERE \citep{2019AJ....157...39G} detections of the HD 115600 disk show a ring of dust at roughly 46 au radius.  \citet{2015ApJ...807L...7C} report a possible (2$\sigma$) stellocentric offset that could be attributed to an eccentric planet \citep{thilliez17a}, but the offset was not confirmed by \citet{2019AJ....157...39G}.  An optical image with HST/STIS over a wider field of view does not detect the disk beyond $\sim$1.5'' (164 au) radius (T. Esposito, priv. comm.) and shows no ring or spiral structures that would be evidence for a dynamical perturbation.

\subsubsection{HD 142315}\label{dis:142315}
HD 142315 (B9V) had very recent encounters with HD 142705 (A0V) at $d_\mathrm{ca}=1.60\pm0.77\,\mathrm{pc}$ and $t_\mathrm{ca}=-0.26_{-0.98}^{+0.62}\,\mathrm{Myr}$ and UCAC4 336-078484 (M0.5) at $d_\mathrm{ca}=1.36_{-0.45}^{+0.50}\,\mathrm{pc}$ and $t_\mathrm{ca}=-0.17_{-0.79}^{+0.46}\,\mathrm{Myr}$; given the uncertainties, these two flyby events are at their closest distances at the present epoch. Figure \ref{fig:each_geometry} (panel e) shows that HD 142705 and UCAC4 336-078484 have very similar posterior distributions in the $d_\mathrm{ca}$-- $t_\mathrm{ca}$ space, suggesting that the two stars share similar space velocities and may be a gravitationally-bound wide binary. These two stars are separated on the sky by 0.20 deg, and have a three-dimensional separation of $1.40_{-0.75}^{+1.36}\,\mathrm{pc}$. At this separation, the escape velocity for the two stars assuming a circular orbit is $0.11_{-0.03}^{+0.05}$\,km\,s$^{-1}$. From their astrometry and radial velocities, we measured a relative velocity of $3.28_{-2.26}^{+3.52}$\,km\,s$^{-1}$. When considering the uncertainties, this is not significantly different from the predicted escape velocity. More precise radial velocity measurements of the two stars are required in order to refine the estimate of their relative velocities. HD 142315 is also predicted to encounter HD 142097 (A5V) in the future at $d_\mathrm{ca} = 1.55_{-0.12}^{+0.19}\,\mathrm{pc}$ and $t_\mathrm{ca} = 0.59_{-0.22}^{+0.35}\,\mathrm{Myr}$.

\subsubsection{2MASS J16035767-2031055 (J1603-2031)}\label{dis:j1603-2031}
We calculated that the K0 star 2MASS J16035767-2031055 had a very recent close approach with 2MASS J16053815-2039469 (M0) at $t_\mathrm{ca}=-0.43_{-0.39}^{+0.17}\,\mathrm{Myr}$ with a close approach distance at $d_\mathrm{ca}=0.52_{-0.16}^{+0.29}\,\mathrm{pc}$. We also identified that the K-star will have another close approach with RX J1603.9-2031B at $t_\mathrm{ca}=0.96_{-1.23}^{+1.80}\,\mathrm{Myr}$ with a close approach distance of $d_\mathrm{ca}=0.71_{-0.48}^{+2.06}\,\mathrm{pc}$. Incorporating the position angle and inclination derived by \cite{2017ApJ...851...85B} based on the ALMA continuum detection, we computed the flyby angle of the two events to be $\theta_\mathrm{ca}=-11.52_{-27.92}^{+29.51}\,\mathrm{deg}$ and $\theta_\mathrm{ca}=-11.65_{-29.84}^{+33.98}\,\mathrm{deg}$, respectively. The flyby angle is basically unconstrained due to the the large ($\sim30\%$) uncertainties in the fitted disk orientation. Additionally, \cite{2020A&A...633A..82G} carries out observations on this star in $H$-band using SPHERE but reports a non-detection of any disk in scattered light.

J1603-2031 (also catalogued as RX J1603.9-2031A) and RX J1603.9-2031B were originally paired by \citet{2000A&A...356..541K}, but their relative velocities are not consistent with them being gravitationally-bound. RX J1603.9-2031B is itself known to be a binary of two stars with similar magnitudes on a 53-year orbit \citep{2018AJ....156..138T}. The RV semi-amplitude of their assumed orbit is approximately 3.6\,km\,s$^{-1}$. While this is larger than the uncertainties on the radial velocity measurement used for the analysis presented here, the measured velocity of the blended spectrum may not be significantly affected by binary motion due to the similar magnitudes of both components. A joint fit of the astrometry presented in \citet{2018AJ....156..138T} and radial velocity measurements where the two stars are spectrally resolved will be needed to determine the systemic velocity, refining the trajectory of this binary relative to J1603-2031.

\subsubsection{2MASS J16042165-2130284 (J1604-2130)}\label{dis:j1604-2130}
\begin{figure*}
    \gridline{\fig{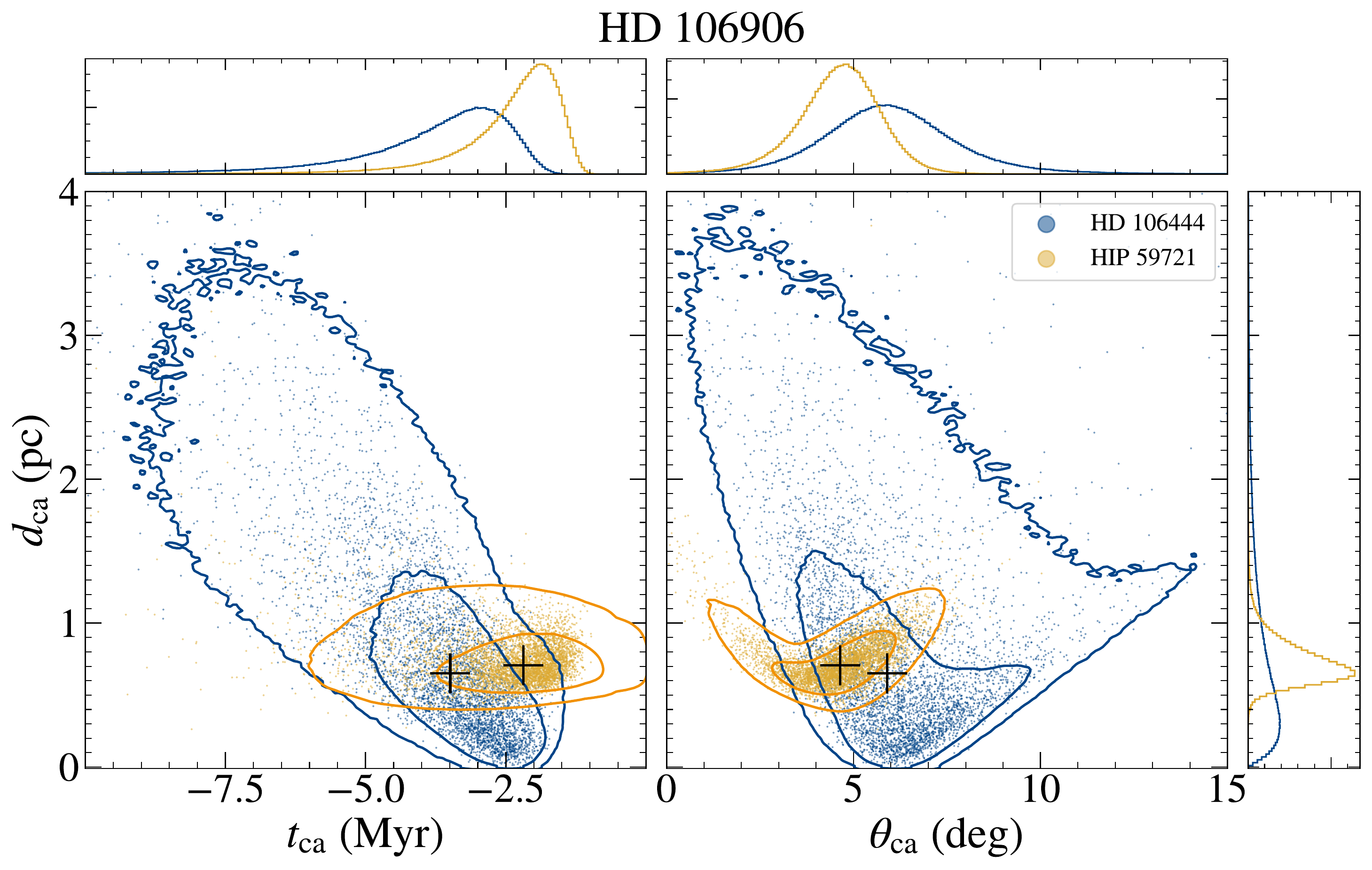}{0.33\textwidth}{(a)} \fig{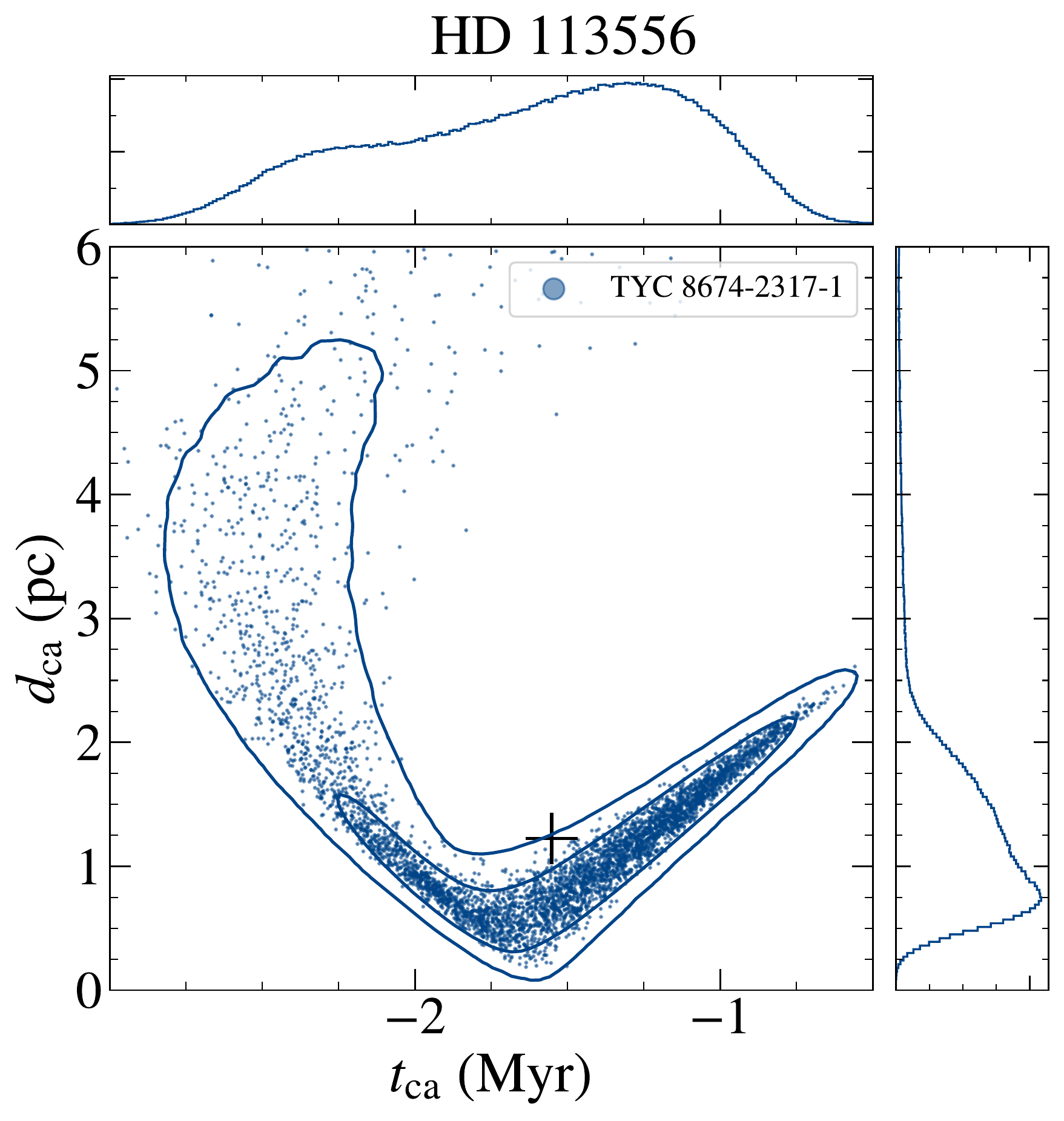}{0.2\textwidth}{(b)}\fig{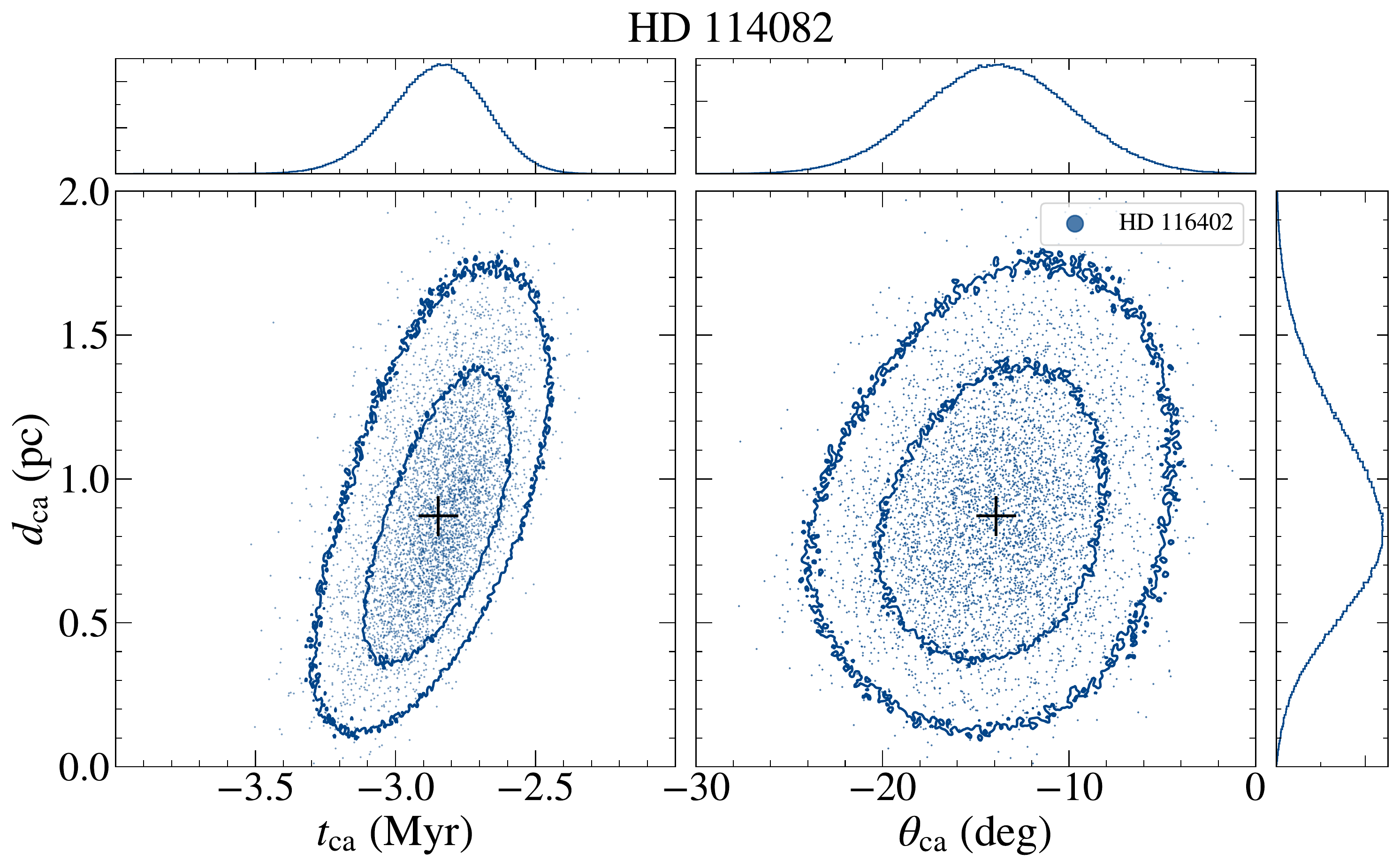}{0.33\textwidth}{(c)}}
    
    \gridline{\fig{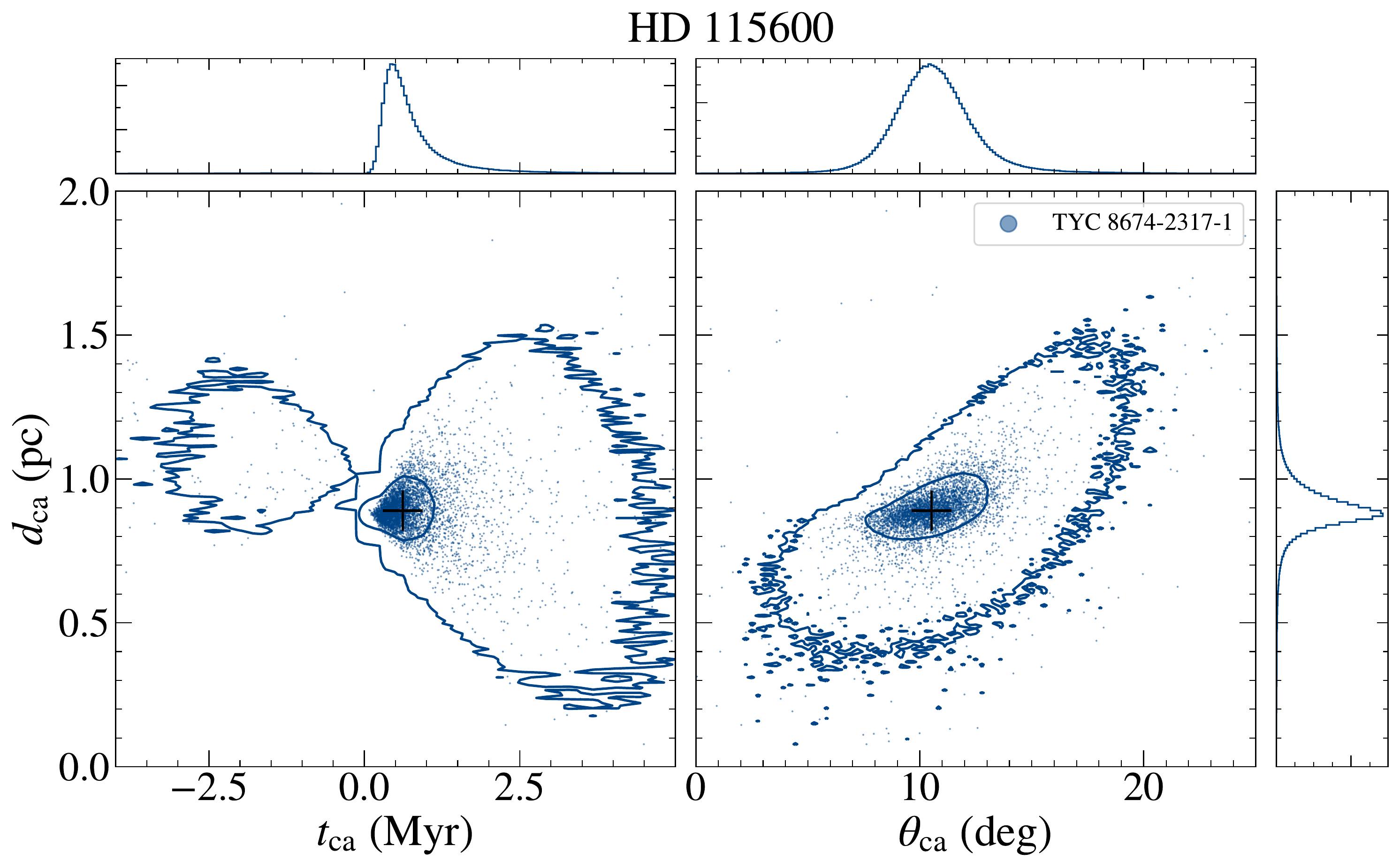}{0.33\textwidth}{(d)}\fig{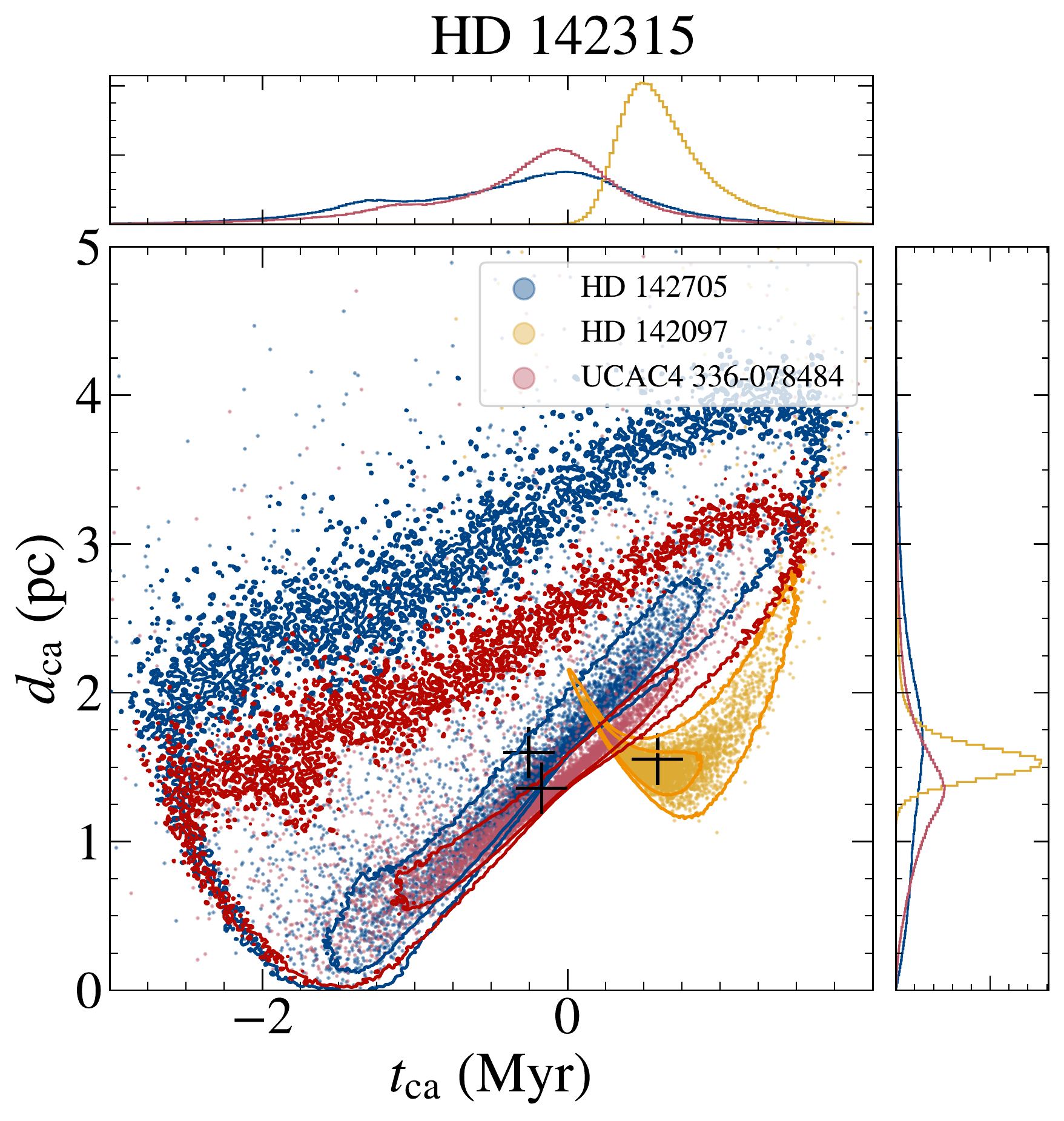}{0.20\textwidth}{(e)}\fig{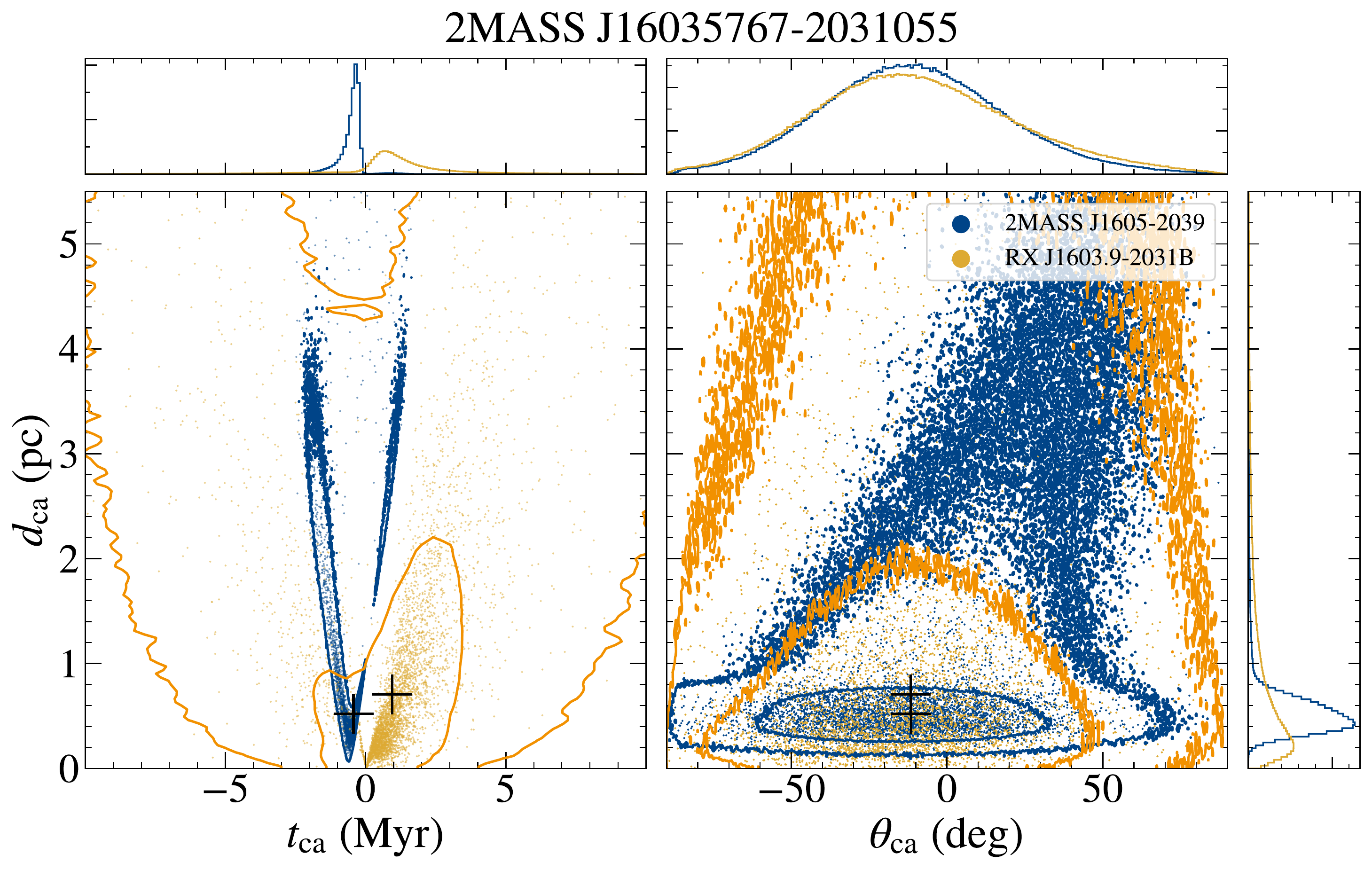}{0.33\textwidth}{(f)}}
    
    \gridline{\fig{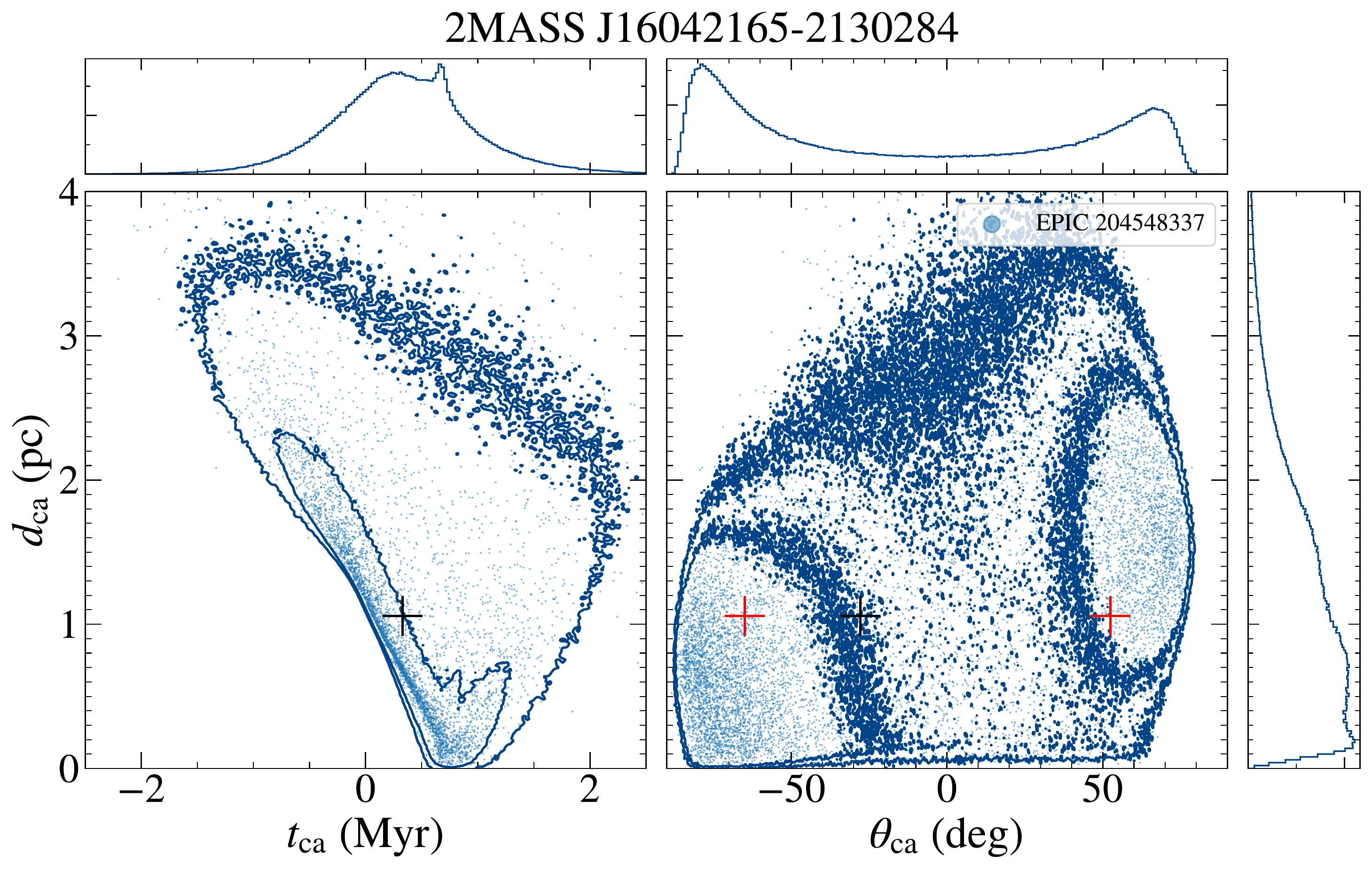}{0.33\textwidth}{(g)}\fig{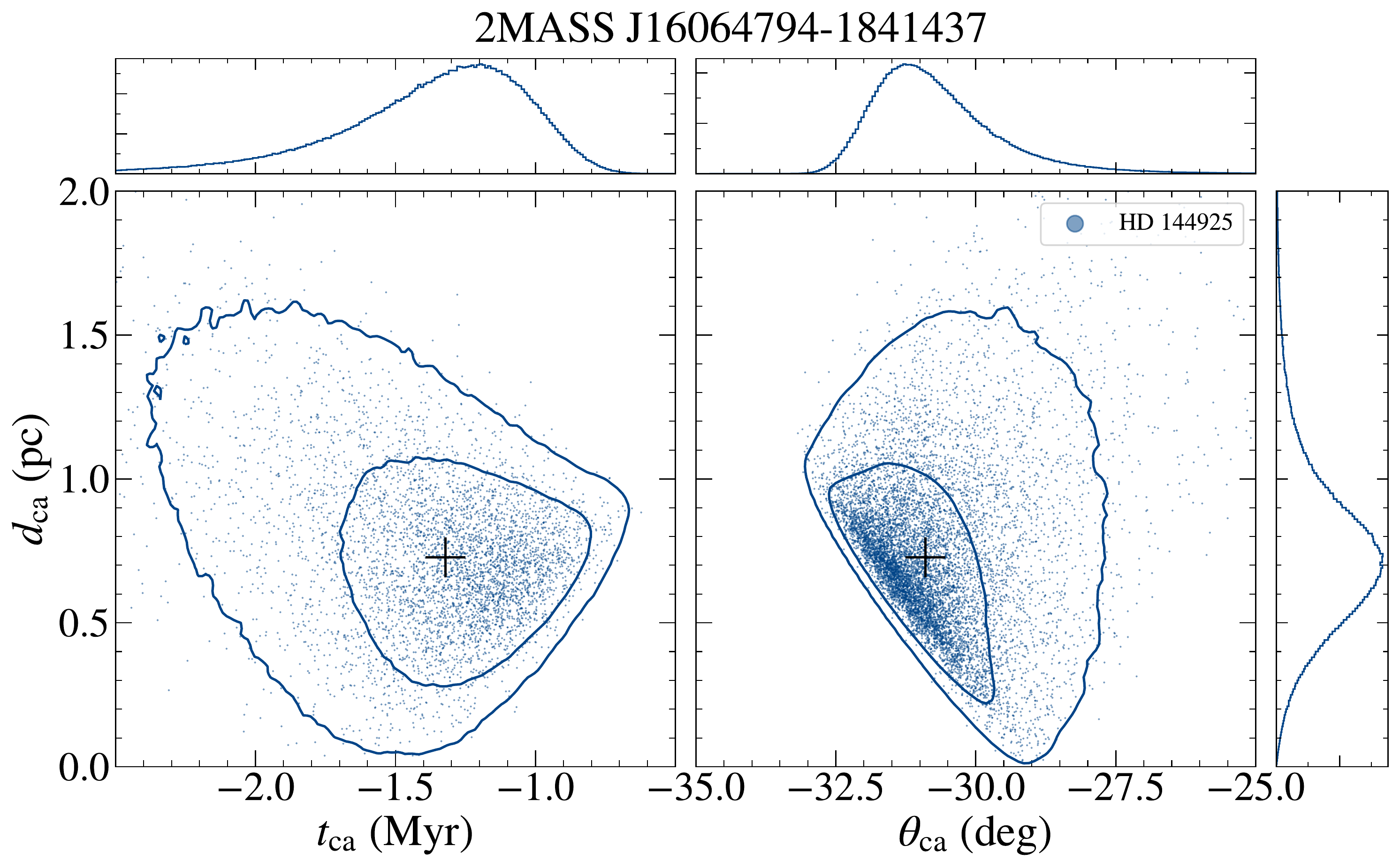}{0.33\textwidth}{(h)}\fig{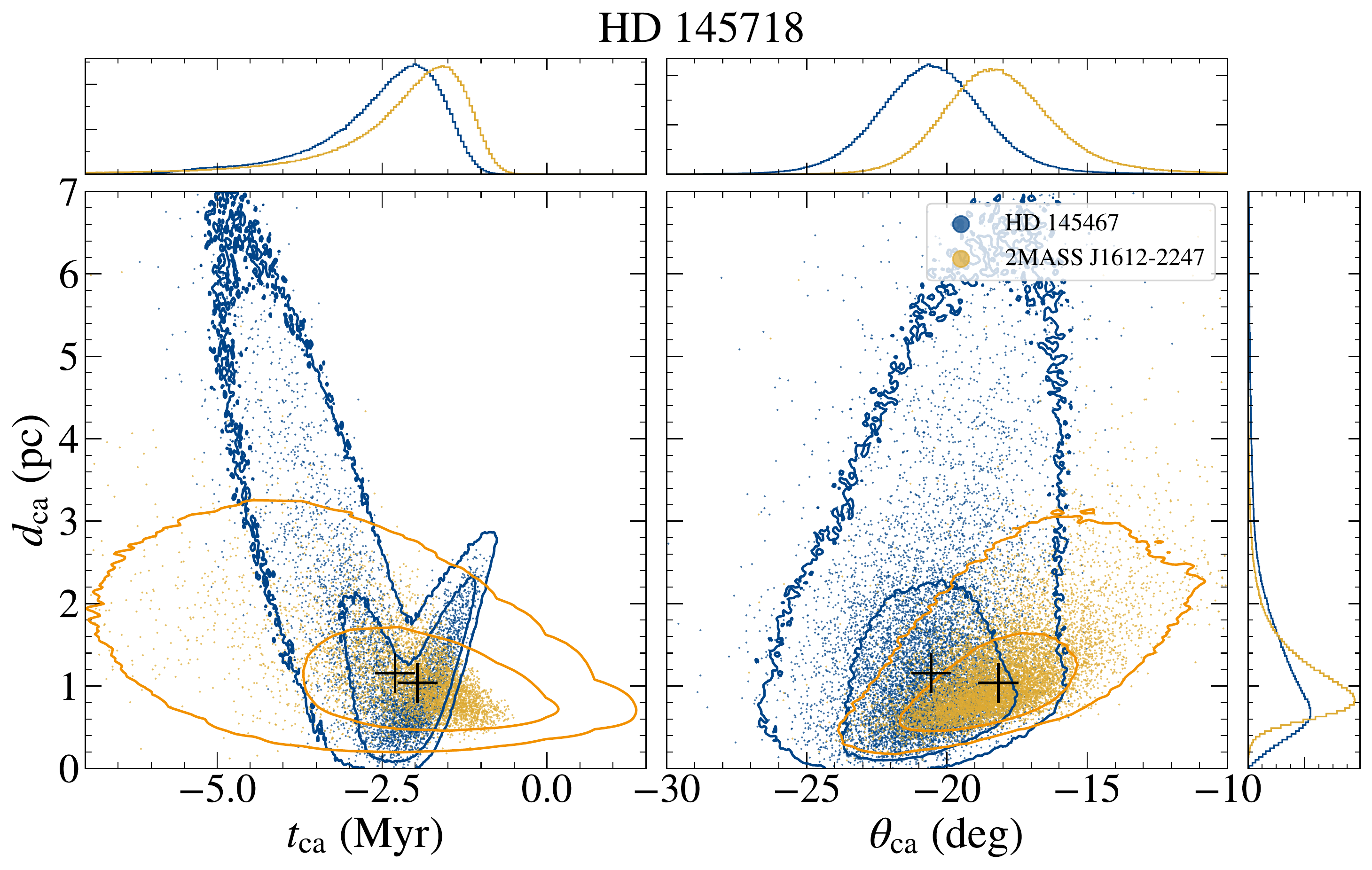}{0.33\textwidth}{(i)}}
    
    \gridline{\fig{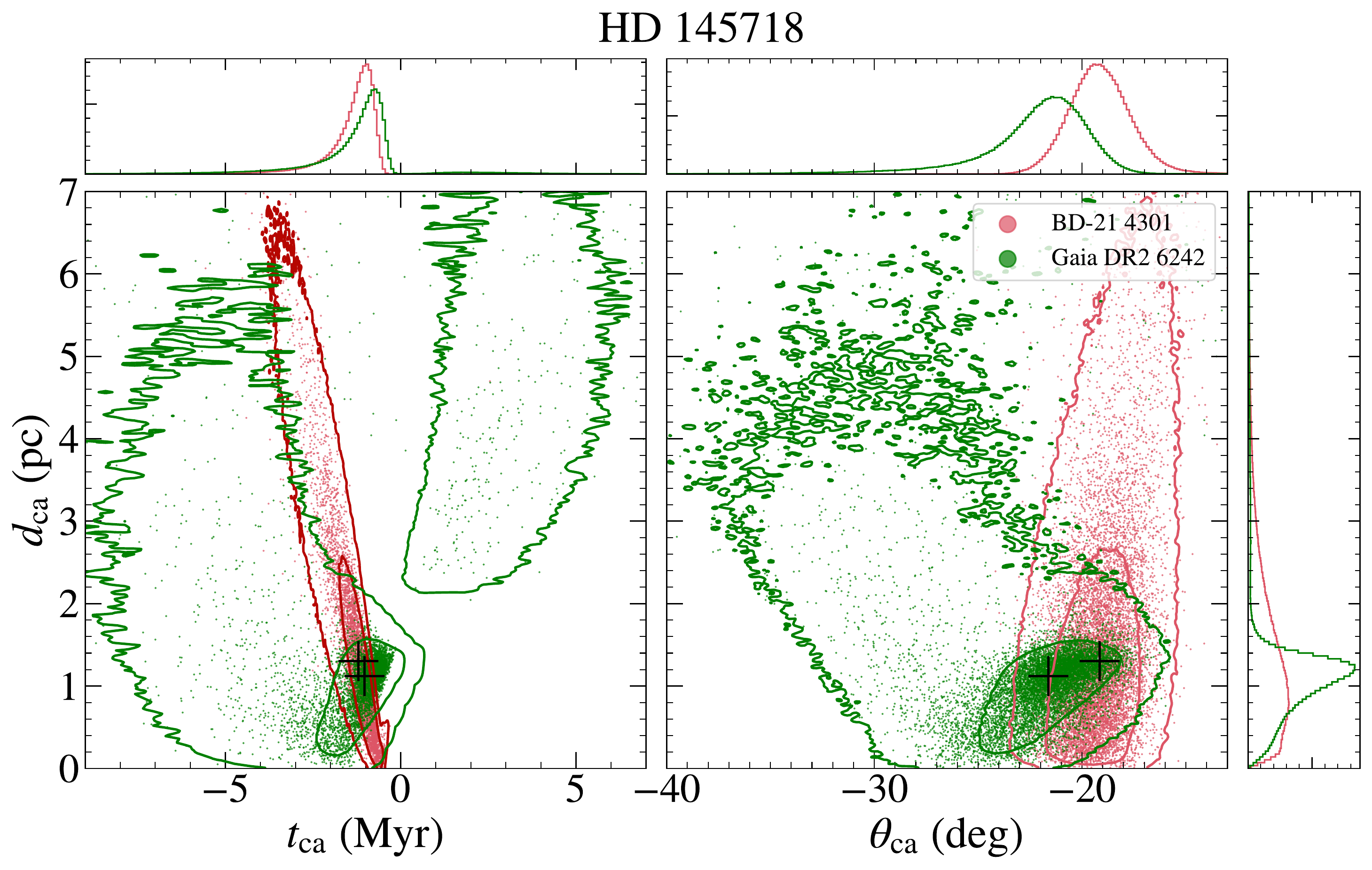}{0.33\textwidth}{(j)}\fig{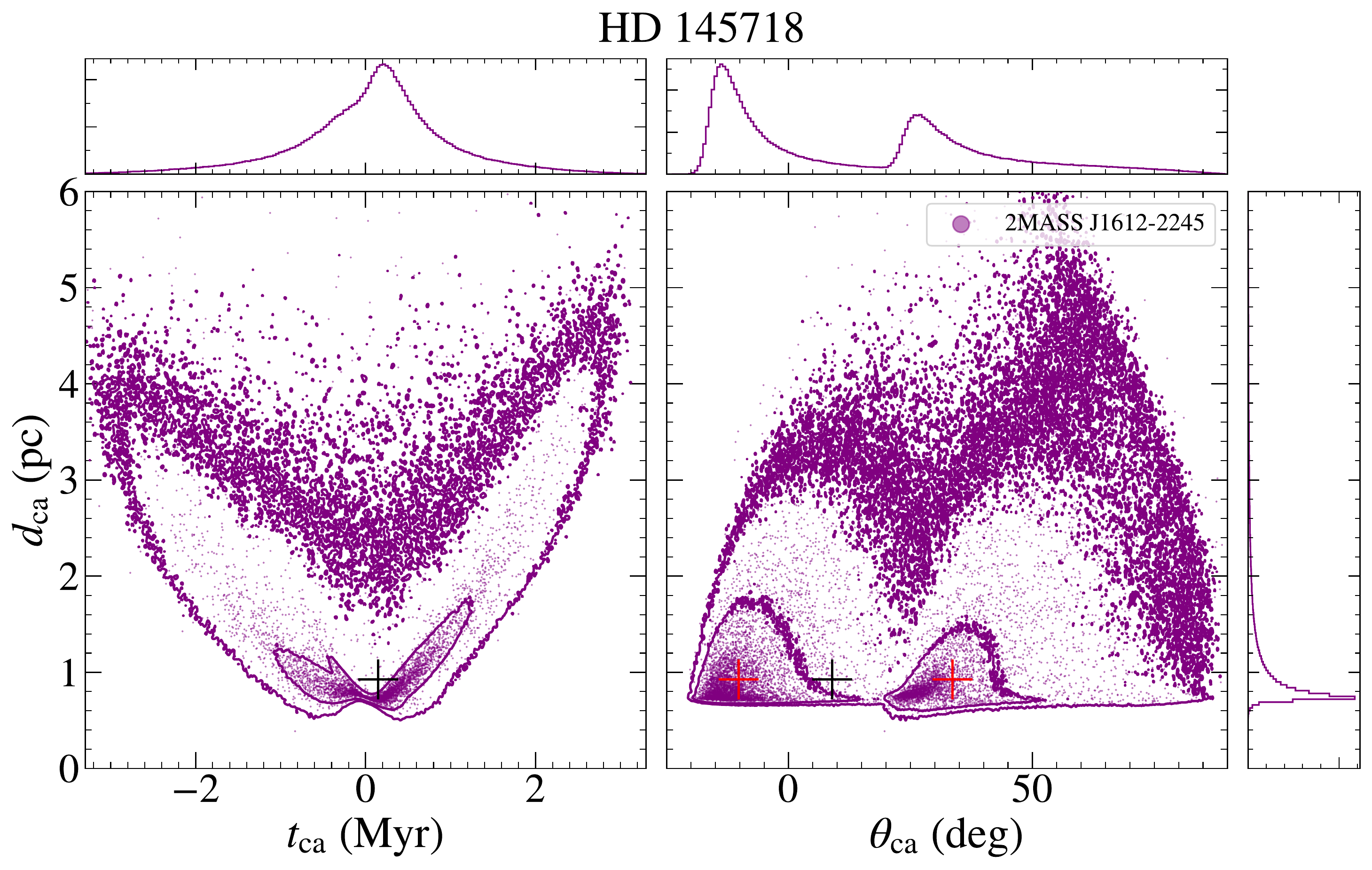}{0.33\textwidth}{(k)}\fig{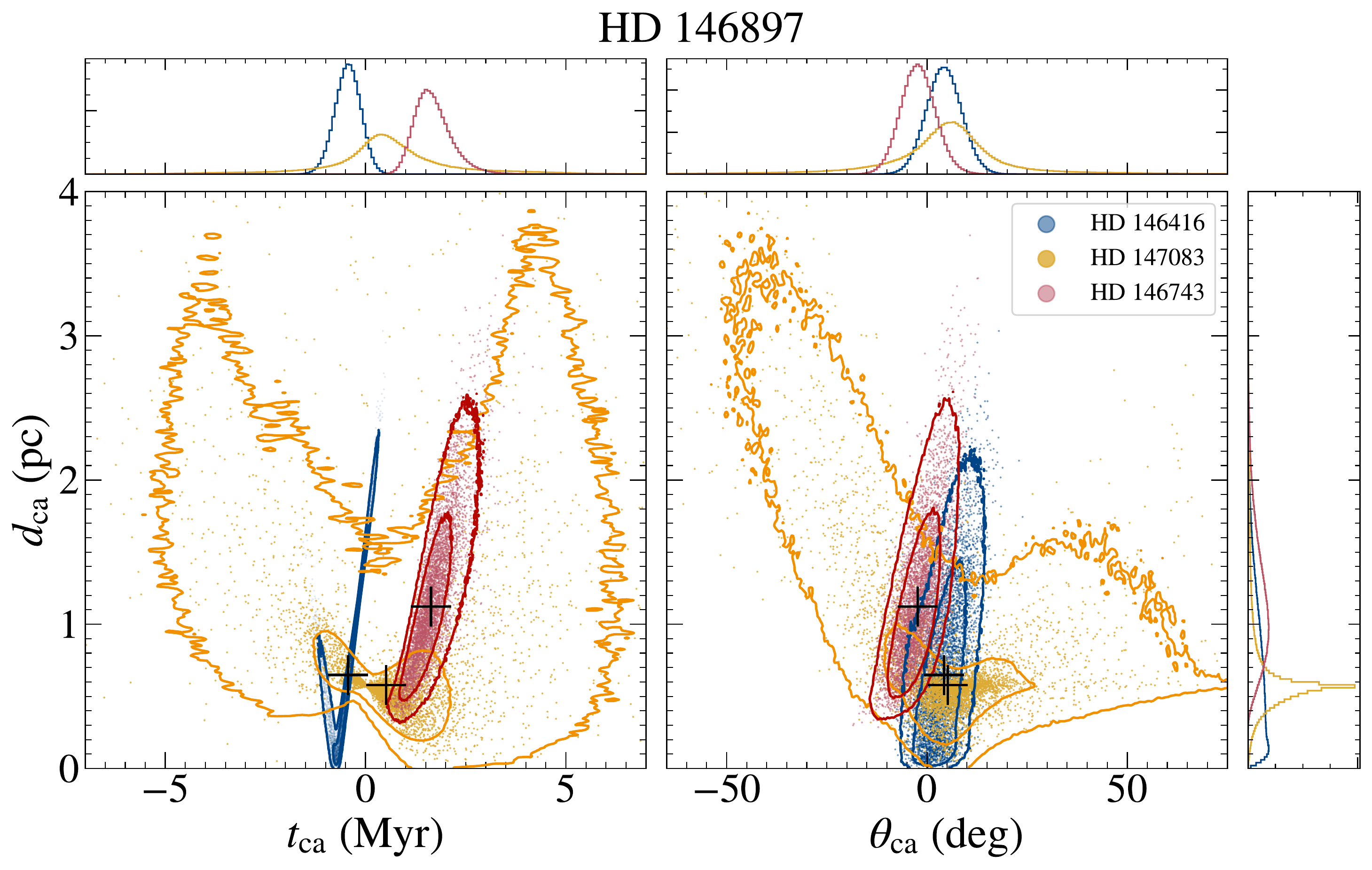}{0.33\textwidth}{(l)}}
    
    \caption{Closest approach distance as a function of closest approach time and  closest approach angle for each of the flyby events involving detected debris disks. The crosses represent the median values of MC outputs; if the posterior distribution is bimodal, the median about each mode is marked by a red cross. The contours represent the $1\sigma$ and $2\sigma$ credible regions.}
    \label{fig:each_geometry}
\end{figure*}

This K2 star will encounter the M1 star EPIC 204548337 (2MASS J16032787-2153155) with $d_{\text{ca}} = 1.22_{-0.53}^{+0.88}\,\text{pc}$ at $t_\mathrm{ca} = 0.33_{-0.59}^{+0.56}\,\mathrm{Myr}$. Given the uncertainties in $t_\mathrm{ca}$ the closest approach epoch is near the present time and the current separation between the two stars is shown in Table \ref{table:geometry}. J1604-2130 hosts one of the largest disks in Upper Scorpius subgroup, appearing as a face-on dust and gas ring between roughly 15 and 300 au radius in thermal emission \citep{mathews12a, zhang14a, 2016ApJ...827..142B, dong17a} and scattered light \citep{mayama12a, pinilla15a, pinilla18a}. The median close approach angle is calculated to be $-27.59_{-47.42}^{+86.39}$ (Table \ref{table:geometry}). However, due to the bimodality of the posterior distribution of the close approach angle as displayed in panel (g) of Fig.\ref{fig:each_geometry}, we divided the distribution at $\theta_\mathrm{ca}=0^\circ$ and calculated the median of each section to be $52.46_{-30.30}^{+15.75}\,\mathrm{deg}$ and $-64.89_{-14.05}^{+34.18}\,\mathrm{deg}$, which suggests that the flyby of EPIC 204548337 is nearly perpendicular to the disk plane. The azimuthal structure of the dust ring has a rotating deficit of scattered light that can be explained as shadowing by optically thick clumps orbiting within an inner dust disk. The existence of a clumpy inner disk misaligned with respect to the outer disk is supported by the observation of aperiodic photometric dips \citep{ansdell16a}. Future work is needed to investigate if EPIC 204548337 plays a dynamical role in misaligning the outer disk relative to the inner disk (e.g., \citealt{kraus20a, ginski21a}).

\subsubsection{2MASS J16064794-1841437 (J1606-1841)}\label{dis:j1606-1841}
J1606-1841 (M0.0e) has an ALMA-resolved disk with $i=55.5^{+0.1}_{-0.1}$ deg and PA=$20^{+1}_{-1}$ deg \citep{ansdell20a}. The bi-lobed thermal emission map suggests a narrow ring-like structure. The flyby star HD 144295 (A0V) came within 0.7 $R_H$ (Table \ref{table:all_luhman_encounters}) at $t_\mathrm{ca} = -1.32_{-0.40}^{+0.26}\,\mathrm{Myr}$. The high relatively velocity $v_\mathrm{rel}$ = 13.46 km/s (Table \ref{table:geometry}) makes it unlikely that the disk was significantly perturbed by the encounter. On the other hand, HD 144295 (HIP 79124) is a triple stellar system \citep{asensio-torres19a, ruane19a} that needs future characterization to improve the determination of the systemic velocity. 

\subsubsection{HD 145718}\label{dis:hd145718}
In Table \ref{table:geometry} this A5 star it is distinguished by having a total of five flybys, all within $\sim$2.5 million years. 
The flyby of 2MASS J16125889-2245202 (M0) has the smallest $d_\mathrm{ca}$ (0.93 pc), corresponding to 0.65 $R_H$  (Table \ref{table:all_luhman_encounters}). \citet{ansdell20a} report that HD 145718 has an ALMA-resolved disk with PA=$1^{+1}_{-1}$ deg and $i=70.4^{+1.2}_{-1.2}$ deg. The inner circumstellar disk was resolved using VLTI GRAVITY in $K$-band, giving PA=$2_{+2}^{-2}$ deg and $i=71.9^{+1.2}_{-1.2}$  \citep{gravity19a}. It was also resolved with VLTI PIONEER in $H$-band \citep{kluska20a}, giving PA=$-3^{+4}_{-4}$ deg and $i=48^{+3}_{-3}$ deg. The main discrepancy is in the determination of disk inclination.  For the purpose of calculating the encounter geometry, we adopt the values given by \citet{ansdell20a}. We note that the posterior distribution of the close approach angle with 2MASS J16125889-2245202 is bimodal---this is likely due to the fact that both stars' RV measurements are not certain about the direction of the line-of-sight motion. Therefore, we divided the distributions at $\theta_\mathrm{ca}=18^\circ$ into two sections and calculated the median of each section to be $33.64_{-7.98}^{+22.34}\,\text{deg}$ and $-10.22_{-4.23}^{+10.67}\,\text{deg}$, respectively. 

\subsubsection{HD 146897}\label{dis:146897}
We find three flybys for HD 146897 (F2/F3V; HIP 79977), two in the recent past or present, and one in the future. HD 146897 had a close encounter with HD 146416 (B9V) at $t_{\text{ca}} = -0.44_{-0.33}^{+0.31}\,\text{Myr}$ and $d_{\text{ca}}=0.65_{-0.43}^{+0.61}$ pc. The flyby is nearly coplanar with the debris disk ($\theta_{\text{ca}}=4.23_{-4.34}^{+4.37}\,\text{deg}$) which was first resolved in scattered light by \citet{thalmann13a}.
\cite{2018AJ....156..279G} measured a brightness asymmetry along the midplane in images obtained with Subaru SCExAO/CHARIS.  This asymmetry is also seen in GPI images \citep{esposito_2020} and extends to $>$6'' (789 au) radius in optical HST/STIS images \citep{kalas20a}.

\par HD 146897 also has a close encounter with HD 147083 (A7III/IV) in the future or present with $t_{\text{ca}}=0.51_{-1.09}^{+1.34}\,\text{Myr}$, $d_{\text{ca}} = 0.58_{-0.13}^{+0.46}\,\text{pc}$. The closest approach angle is $\theta_{\text{ca}} = 5.19_{-12.69}^{+10.26}\,\text{deg}$. The error bar on $\theta_{\text{ca}}$ is large compared to the other flyby events because of the large uncertainty in the radial velocity of the flyby star. 

\par Lastly, HD 146897 will possibly encounter HD 146743 (F3V) at $t_{\text{ca}} = 1.63_{-0.39}^{+0.50}\,\text{Myr}$ and $d_{\text{ca}} = 1.12_{-0.38}^{+0.53}\,\text{pc}$. However, it is unlikely this encounter will affect the current disk morphology given the large closest approach distance.  

\section{Potential Encounters in Upper Scorpius Subgroup}\label{sec:usco_encounters}
\begin{figure}
    \includegraphics[width=\columnwidth]{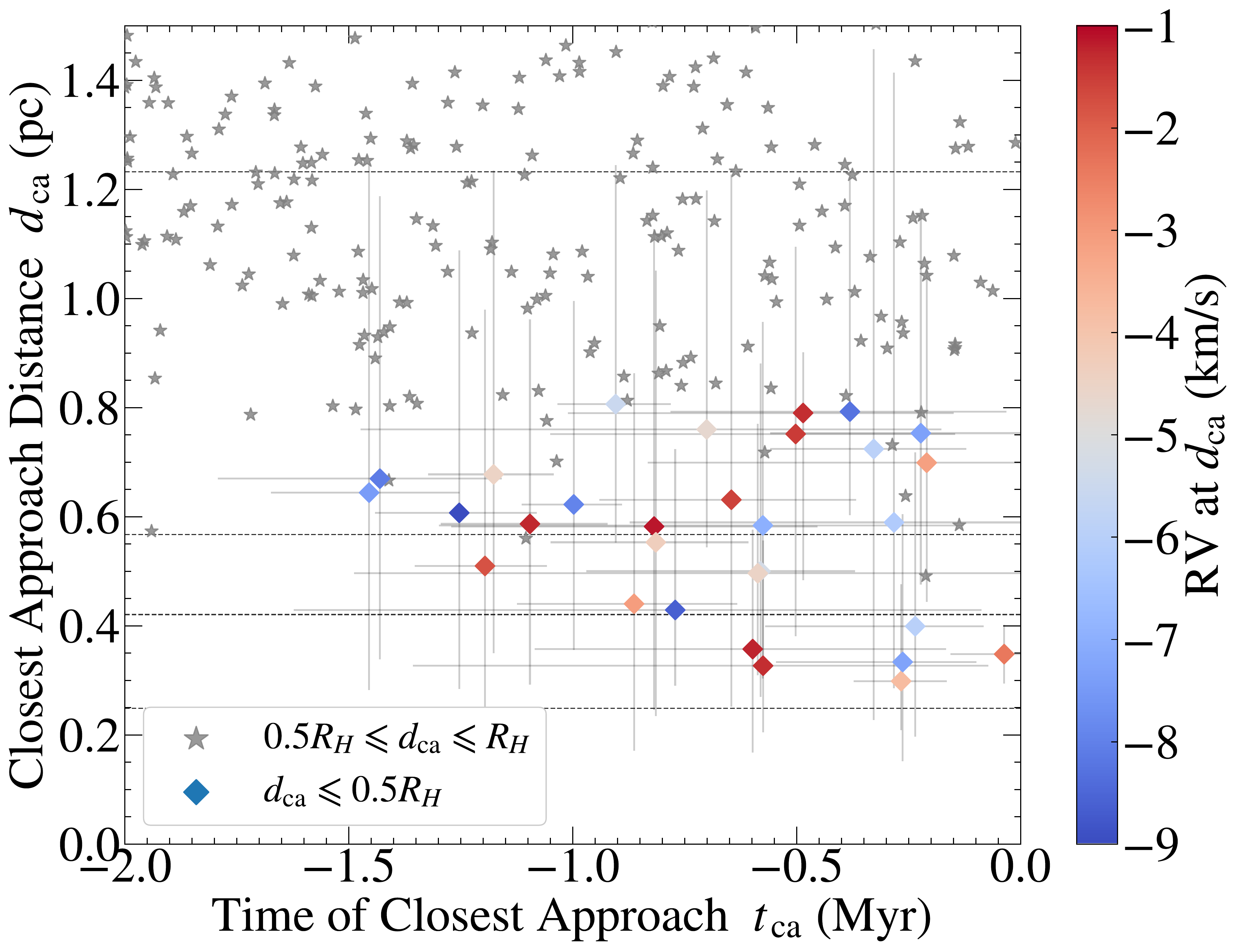}
    \caption{The potential close approaches that would have happened within the USCO subgroup by simulating the effect of perfectly knowing the RV of stars without RV measurement. The dotted lines mark the 1st-, 5th-, 10th-, and 50th-percentiles of the distribution of the separation of stars within the sample. }
    \label{fig:single_rv_dt}
\end{figure}
There are $1341$ stars in the USCO sample that do not have RV measurement. We also attempted to identify potential close approaches between these stars and those that have RV measurements. While we cannot confidently determine which of these stars will have experienced a flyby event in the past, we can identify candidate encounters to motivate future radial velocity measurements of these stars to better understand their kinematics. For each pair of stars (let the one with RV measurement be star A and the other be star B), we sampled the 6D space position and motion of star A derived from \textit{Gaia} astrometry and radial velocities, and the five known astrometric parameters of star B with $10^5$ trials using the method in Section \ref{sec:flyby_identification}. For star B the radial velocity was fixed at a constant value, allowing us to identify candidate flyby events assuming this velocity. We repeated this exercise for radial velocities consistent with the distribution of other Upper Sco members ($-5\pm4$\,km\,s$^{-1}$; \citep{gagne_2018}), changing the radial velocity from $-21$\,km\,s$^{-1}$ to $11$\,km\,s$^{-1}$ ($-4\sigma$ to $+4\sigma$) in steps of 0.01\,km\,s$^{-1}$.

Flyby events identified from this analysis that had $t_{\rm CA}<0$\,Myr and where the radial velocity of star B was within the $1\sigma$ range of the distribution of Upper Sco members were selected as candidate flyby events within the USCO sample. Overall, we identified 30 potential close encounters and 224 potential encounters. The distribution of these candidate events in terms of closest approach time and distance is shown in Figure \ref{fig:single_rv_dt}, and a complete listing is given in Table \ref{table:single_rv_flyby}.

Of these candidate encounters, four involve three stars with a measured IR excess in \cite{cotten2016}; HD 144587, HD 146069, and HD 145631 that has two candidate encounters. The paucity of excess stars within this sample is not surprising; many of these stars have only recently been identified as members of the Upper Sco subgroup. Of these three, only HD 144587 has been observed by a high-contrast imager to search for polarized emission from circumstellar material, but no detection was reported \citep{2017AJ....153..106U}. All of the stars have been the subject of adaptive optics searches for stellar companions. HD 145631 was resolved as a hierarchical triple system consisting of an F-type primary and at 191\,au two K-type companions with a projected separation of 7.3\,au \citep{2014ApJ...785...47L}. It is not known what effect these companions are having on the morphology of the disk as it has not yet been spatially resolved.

\section{Conclusion}\label{sec:conclusion}
Using astrometry from \textit{Gaia} and radial velocities from a combination of literature sources, we searched for close stellar flybys within the Scorpius-Centaurus OB2 association. We calculated the closest approach distance and the time at which the close approaches take place for two samples of stars. The first consisted of 462 stars spanning all three sub-groups of Sco-Cen whose membership were assessed primarily from {\textit Hipparcos} astrometry (e.g., \citealp{1999AJ....117..354D}), while the second contained 219 stars within the Upper Sco subgroup identified from a more recent analysis of \textit{Gaia DR2} astrometry \citep{luhman20}. We identified 12 past close encounters (defined as one star entering the half hill sphere radius of another), and 12 future close encounters amongst the first sample, and 37 more distant encounters. Amongst the Upper Sco sample, we identified 36 past close encounters, 18 future close encounters, and 106 more distant encounters. We also searched for candidate flyby events involving the Upper Sco sample members that do not yet have radial velocity measurements. We identified 41 candidate close encounters given simulated radial velocities consistent with the measured distribution of other Upper Sco members \citep{gagne_2018}.

Motivated by theoretical work exploring the interaction between stellar flybys and the evolution of exoplanetary system architectures, we cross-matched the flyby events identified within this work with catalogues of stars within Sco-Cen that have measured infrared-excesses \citep{cotten2016}, indicative of the presence of circumstellar material. For the subset of stars with an infrared excess we searched the literature for spatially resolved detections of these disks from which their geometry can be measured. A total of 10 stars with at least one identified flyby event have debris disks that have been resolved either in scattered light from ground or space-based instrumentation, or detected at sub-millimeter wavelengths with ALMA. In one case, for HD 106906, the disk exhibits a strong brightness asymmetry \citep{kalas_2015}, but it is likely that this is caused by the orbiting planetary-mass companion \citep{2014ApJ...780L...4B,2021AJ....161...22N} rather than a direct interaction between the disk and the two flyby stars. For the remaining targets the existing evidence does not show significantly disturbed morphologies that could be caused by a flyby of another star.

Our incomplete census of the full Sco-Cen association limited the analysis presented here to the more massive stars ($>1$\,M$_{\odot}$) for two of the three subgroups. A more exhaustive search for flyby events can be undertaken when the membership of the three subgroups are re-assessed using data from current and future \textit{Gaia} data releases. Another significant limitation is the precision of the radial velocity measurements for many of the stars within our sample; a large uncertainty in the radial velocity at the current epoch significantly hampers our ability to predict the relative positions and velocities of pairs of stars within the cluster at earlier times. Future \textit{Gaia} data releases will contain radial velocity measurements for many more of the cluster members than used in this study, but at a relatively low precision ($\sim$1\,km\,s$^{-1}$). In order to accurately trace back the motions of the cluster members to confidently determine which pairs of stars experienced a close encounter in the past, precision radial velocities are needed in conjunction with the precision astrometry provided by \textit{Gaia}. With these data, we will be able investigate more interesting and stronger dynamical interactions between flyby stars and debris disks to provide empirical evidence of the impact of stellar flybys on the evolution of planetary systems.

\acknowledgments
We thank the anonymous referees for their comments that helped to improve the quality of this manuscript. The first author also thanks Eugene Chiang, Fanghui Wan, and Tianqi Wei for the valuable discussions and/or support regarding the work. This work is supported by NSF AST-1518332, NASA NNX15AC89G and NNX15AD95G/NEXSS. This work benefited from NASA's Nexus for Exoplanet System Science (NEXSS) research coordination network sponsored by NASA's Science Mission Directorate. This research has made use of the SIMBAD database and the VizieR catalogue access tool, both operated at the CDS, Strasbourg, France. This work has made use of data from the European Space Agency (ESA) mission {\it Gaia} (\url{https://www.cosmos.esa.int/gaia}), processed by the {\it Gaia} Data Processing and Analysis Consortium (DPAC, \url{https://www.cosmos.esa.int/web/gaia/dpac/consortium}). Funding for the DPAC has been provided by national institutions, in particular the institutions participating in the {\it Gaia} Multilateral Agreement.

\bibliography{flyby}{}
\bibliographystyle{aasjournal}

\appendix

\section{Tables of the two samples and all identified events}

\startlongtable
\begin{deluxetable*}{cccccccccc}
\tablecaption{RV measurements of Sco-Cen sample \label{table:462scocen}}
\tablehead{\colhead{Target} & \colhead{\textit{Gaia} DR2 identifier} & \colhead{SpT} & \colhead{Mass} & \colhead{Subgroup} & \colhead{\textit{Gaia} RV} & \colhead{\textit{Gaia} RV error} & \colhead{RV} & \colhead{RV error} & \colhead{ref.} \\
\colhead{} & \colhead{} & \colhead{} & \colhead{(M$_\odot$)} & \colhead{} & \colhead{(km/s)} & \colhead{(km/s)} & \colhead{(km/s)} & \colhead{(km/s)} & \colhead{}
}
\startdata
HD 103823 & 3463980463840622848 & G2/3V & 1.256 & LCC & $19.3$ & 0.28 & $19.5$ & 2.3 & 14 \\
\vdots & \vdots & \vdots & \vdots & \vdots & \vdots & \vdots & \vdots & \vdots & \vdots\\
\enddata
\tablerefs{(1) \citealp{1965MNRAS.129..191T}; (2)\citealp{kk_1973}; (3) \citealp{1967IAUS...30...57E}; (4) \citealp{1987ApJS...64..487L}; (5) \citealp{mgc1991}; (6) \citealp{levato_1996};
(7) \citealp{1996Obs...116...85S};
(8) \citealp{torres_2006};
(9) \citealp{2006AstL...32..759G};
(10) \citealp{2007AN....328..889K};
(11) \citealp{2011ApJ...738..122C};
(12) \citealp{2012AJ....144....8S};
(13) \citealp{2012ApJ...745...56D};
(14) \citealp{2013AJ....146..134K};
(15) \citealp{desidera_2015};
(16) \citealp{2015MNRAS.448.3784B};
(17) \citealp{2017ApJ...840...87R};
(18) \citealp{2018AJ....156..194T};
(19) \citealp{gaia2018};
(20) \citealp{derosa_kalas}.
}
\tablecomments{The full table will be available in the online version of the paper and on Vizier.}
\end{deluxetable*}
\startlongtable
\begin{deluxetable*}{ccccccccc}
\tablecaption{RV measurements of USCO sample \label{table:usco_sample}}
\tablehead{\colhead{Target} & \colhead{\textit{Gaia} DR2 identifier} & \colhead{SpT} & \colhead{Mass} & \colhead{\textit{Gaia} RV} & \colhead{\textit{Gaia} RV error} & \colhead{RV} & \colhead{RV error} & \colhead{ref.} \\
\colhead{} & \colhead{} & \colhead{} & \colhead{(M$_\odot$)} & \colhead{(km/s)} & \colhead{(km/s)} & \colhead{(km/s)} & \colhead{(km/s)} & \colhead{}
}
\startdata
2MASS J15401546-1620577 & 6261607772589226496 & \nodata & 12.971 & -12.12 & 4.24 & -12.12 & 4.24 & 8 \\
\vdots & \vdots & \vdots & \vdots & \vdots & \vdots & \vdots & \vdots & \vdots \\
\enddata
\tablerefs{(1) \citealp{1987ApJS...64..487L}; (2) \citealp{mgc1991}; (3) \cite{dfm1995}; (4) \citealp{2006AstL...32..759G}; (5) \citealp{2007AN....328..889K}; (6) \citealp{2011ApJ...738..122C}; (7) \citealp{2012ApJ...745...56D}; (8) \citealp{gaia2018}
}
\tablecomments{The full table will be available in the online version of the paper and on Vizier}
\end{deluxetable*}

\begin{longrotatetable}
\begin{deluxetable*}{cccccccc}
\tablecaption{Events identified from Sco-Cen sample\label{table:all_scocen_encounters}}
\tablehead{\colhead{Star 1} & \colhead{Star 2} & 
\colhead{$t_{\text{ca}}$ (Myr)} &\colhead{$d_{\text{ca}}$ (pc)} &
\colhead{$d_{\text{ca}}/R_\text{H,T}$} &
\colhead{$d_{\text{ca}}/R_\text{H,F}$} &
 \colhead{Category} & \colhead{IR Excess?}}
\startdata
2MASS J12143187-5110156 & CD-50 6815 & $0.02_{-0.30}^{+0.30}$ & $0.18_{-0.14}^{+0.40}$ & $0.17_{-0.12}^{+0.36}$ & $0.14_{-0.11}^{+0.31}$ & TF0.5 & \nodata \\
CD-48 8201 & HD 116650 & $-0.21_{-0.09}^{+0.07}$ & $0.51_{-0.28}^{+0.24}$ & $0.42_{-0.23}^{+0.20}$ & $0.39_{-0.22}^{+0.19}$ & TF0.5 & \nodata \\
HD 104231 & HD 104125 & $-0.61_{-0.32}^{+0.26}$ & $0.24_{-0.11}^{+0.28}$ & $0.18_{-0.09}^{+0.21}$ & $0.16_{-0.08}^{+0.18}$ & TF0.5 & T \\
HD 106444 & HIP 59721 & $-0.09_{-0.37}^{+0.32}$ & $0.08_{-0.05}^{+0.16}$ & $0.06_{-0.04}^{+0.12}$ & $0.06_{-0.04}^{+0.13}$ & TF0.5 & \nodata \\
HD 106906 & HD 106444 & $-3.49_{-1.75}^{+0.90}$ & $0.65_{-0.40}^{+0.92}$ & $0.48_{-0.30}^{+0.68}$ & $0.48_{-0.30}^{+0.69}$ & TF0.5 & T \\
HD 107821 & TYC 8983-854-1 & $0.42_{-0.16}^{+0.22}$ & $0.56_{-0.11}^{+0.20}$ & $0.37_{-0.07}^{+0.13}$ & $0.45_{-0.09}^{+0.16}$ & TF0.5 & \nodata \\
HD 110634 & 2MASS J12454884-5410583 & $-0.94_{-0.19}^{+0.14}$ & $0.62_{-0.17}^{+0.14}$ & $0.45_{-0.12}^{+0.10}$ & $0.49_{-0.13}^{+0.11}$ & TF0.5 & T \\
HD 145631 & HD 145502 & $0.34_{-0.48}^{+0.49}$ & $0.64_{-0.20}^{+0.46}$ & $0.39_{-0.12}^{+0.28}$ & $0.27_{-0.09}^{+0.20}$ & TF0.5 & T \\
\enddata
\tablecomments{The letters ``T" and ``F" in the table stand for ``target star" and ``flyby star", respectively. In the ``Category" column, ``T" and ``F" denote encounters with $0.5 < d_\text{ca}/R_\text{H,T}<1$ and $0.5 < d_\text{ca}/R_\text{H,F}<1$, respectively; ``TF" denotes encounters satisfying both previous conditions. Those ending with ``0.5" denenote the fact that $d_\text{ca}/R_\text{H} <= 0.5$, where $R_\text{H}$ is the Hill radius of target star (T), flyby star (F), or both (TF). Similarly, in the ``IR Excess" column, ``T" or ``F" denotes that the target star or the flyby star has IR excess detected, respectively; ``TF" denotes that both stars have IR excess detected, according to \citealt{cotten2016}.}
\end{deluxetable*}
\end{longrotatetable}

\begin{longrotatetable}
\begin{deluxetable*}{cccccccc}
\tablecaption{Events identified from USCO sample\label{table:all_luhman_encounters}}
\tablehead{\colhead{Star 1} & \colhead{Star 2} & 
\colhead{$t_{\text{ca}}$ (Myr)} &\colhead{$d_{\text{ca}}$ (pc)} &
\colhead{$d_{\text{ca}}/R_\text{H,T}$} &
\colhead{$d_{\text{ca}}/R_\text{H,F}$} &
 \colhead{Category} & \colhead{IR Excess?}
}
\startdata
Gaia DR2 6041813096094711424 & Gaia DR2 6041813100399783552 & $0.09_{-0.27}^{+0.29}$ & $0.16_{-0.12}^{+0.41}$ & $0.17_{-0.12}^{+0.43}$ & $0.15_{-0.12}^{+0.40}$ & TF0.5 & \nodata \\
CD-26 11326 & $[$SMH2016$]$ NV6 & $-0.08_{-0.35}^{+0.30}$ & $0.54_{-0.16}^{+0.31}$ & $0.39_{-0.12}^{+0.22}$ & $0.48_{-0.14}^{+0.28}$ & TF0.5 & \nodata \\
2MASS J16210010-2536031 & 2MASS J16211379-2533314 & $-0.03_{-0.16}^{+0.16}$ & $0.30_{-0.14}^{+0.49}$ & $0.24_{-0.11}^{+0.39}$ & $0.25_{-0.12}^{+0.41}$ & TF0.5 & \nodata \\
HD 146001 & 2MASS J16145546-2527597 & $0.25_{-0.43}^{+0.55}$ & $0.46_{-0.32}^{+0.80}$ & $0.26_{-0.18}^{+0.45}$ & $0.34_{-0.23}^{+0.59}$ & TF0.5 & \nodata \\
HD 147196 & 2MASS J16211450-2342200 & $-0.04_{-0.31}^{+0.30}$ & $0.43_{-0.30}^{+0.77}$ & $0.23_{-0.16}^{+0.41}$ & $0.38_{-0.26}^{+0.67}$ & TF0.5 & \nodata \\
2MASS J16053815-2039469 & 2MASS J16035767-2031055 & $-0.43_{-0.39}^{+0.17}$ & $0.52_{-0.16}^{+0.29}$ & $0.49_{-0.14}^{+0.27}$ & $0.43_{-0.13}^{+0.24}$ & TF0.5 & \nodata \\
UCAC4 354-076591 & 2MASS J16111742-1918285 & $0.19_{-0.98}^{+1.11}$ & $0.47_{-0.10}^{+0.25}$ & $0.44_{-0.09}^{+0.24}$ & $0.45_{-0.10}^{+0.24}$ & TF0.5 & \nodata \\
UCAC4 354-076591 & Wa Oph 2 & $-0.05_{-0.14}^{+0.15}$ & $0.46_{-0.18}^{+0.50}$ & $0.43_{-0.17}^{+0.47}$ & $0.36_{-0.14}^{+0.39}$ & TF0.5 & \nodata \\
\enddata
\tablecomments{The notations in this table are the same as in Table \ref{table:all_scocen_encounters}. The target and flyby stars are presented using the \textit{Gaia} DR2 identifiers. Only Gaia DR2 6237422850406493952 (HD 142315) has been observed by \cite{esposito_2020} but no disk was detected.}
\end{deluxetable*}
\end{longrotatetable}

\begin{longrotatetable}
\begin{deluxetable*}{cccccc}
\tablecaption{Events involving candidates and field stars with $d_{\text{ca}} < 1$\,pc and $v_\mathrm{rel}< 10$\,km\,s$^{-1}$ \label{table:candidate_field}}
\tablehead{\colhead{Star 1} & \colhead{Star 2} & 
\colhead{$t_{\text{ca}}$ (Myr)} &\colhead{$d_{\text{ca}}$ (pc)} &
\colhead{$v_\mathrm{rel}$ (km\,s$^{-1}$)} &
 \colhead{Category}}
\startdata
Gaia DR2 3463980463840622848 & Gaia DR2 3463980459544094208 & $0.01_{-0.34}^{+0.33}$ & $0.39_{-0.28}^{+0.46}$ & $0.80_{-0.10}^{+0.24}$ & F \\
Gaia DR2 3481965141177021568 & Gaia DR2 3481965995873045888 & $-0.19_{-0.36}^{+0.53}$ & $0.25_{-0.18}^{+0.59}$ & $4.80_{-3.15}^{+4.83}$ & F \\
Gaia DR2 4126469681670065408 & Gaia DR2 4126469685970383360 & $-0.53_{-0.31}^{+0.27}$ & $0.42_{-0.20}^{+0.25}$ & $3.96_{-0.82}^{+0.83}$ & C \\
Gaia DR2 5334680613863027200 & Gaia DR2 5334681571583520512 & $-0.43_{-0.63}^{+0.51}$ & $0.61_{-0.32}^{+0.89}$ & $7.66_{-2.28}^{+2.29}$ & C \\
Gaia DR2 5341470132513014272 & Gaia DR2 5341856262911551872 & $0.29_{-1.36}^{+0.57}$ & $0.81_{-0.31}^{+0.28}$ & $1.91_{-1.25}^{+1.98}$ & C \\
Gaia DR2 5343524947923387520 & Gaia DR2 5342085785963213184 & $-1.10_{-0.75}^{+0.37}$ & $0.97_{-0.30}^{+0.30}$ & $5.93_{-2.33}^{+2.33}$ & C \\
Gaia DR2 5343610331876174336 & Gaia DR2 5343603288120259072 & $-0.18_{-0.26}^{+0.32}$ & $0.79_{-0.22}^{+0.92}$ & $4.39_{-1.90}^{+3.71}$ & C \\
Gaia DR2 5359955053246566144 & Gaia DR2 5347989239999386240 & $0.41_{-0.12}^{+0.23}$ & $0.76_{-0.04}^{+0.08}$ & $3.08_{-1.15}^{+1.16}$ & F \\
\enddata
\tablecomments{Star 1 are stars from the Sco-Cen sample and Star 2 are stars from the candidate and field star population. All identifiers are from \textit{Gaia} DR2. The category column uses ``C" and ``F" to represent ScoCen-candidate encounters and ScoCen-field encounters.}
\end{deluxetable*}
\end{longrotatetable}

\begin{longrotatetable}
\begin{deluxetable*}{ccccccccc}
\tablecaption{Simulated Potential Flyby Events from USCO Sample \label{table:single_rv_flyby}}
\tablehead{\colhead{Target Star} & \colhead{Flyby Star} & \colhead{$\mathrm{RV}_\text{sim}$} &
\colhead{$t_{\text{ca}}$} &\colhead{$d_{\text{ca}}$} &
\colhead{$d_{\text{ca}}/R_\text{H,T}$} &
\colhead{$d_{\text{ca}}/R_\text{H,F}$} &
 \colhead{Category} & \colhead{IR Excess?}\\
\colhead{} & \colhead{} & \colhead{(km/s)} &
\colhead{(Myr)} &\colhead{(pc)} &
\colhead{} & \colhead{} & \colhead{} & \colhead{} 
}
\startdata
HD 146285 & 2MASS J16172615-2450592 & $-3.76$ & $-0.27_{-0.11}^{+0.10}$ & $0.30_{-0.09}^{+0.18}$ & $0.17_{-0.05}^{+0.10}$ & $0.34_{-0.10}^{+0.20}$ & TF0.5 & \nodata \\
HD 145333 & 2MASS J16140211-2301021 & $-1.57$ & $-0.65_{-0.29}^{+0.28}$ & $0.63_{-0.38}^{+0.60}$ & $0.44_{-0.26}^{+0.41}$ & $0.47_{-0.28}^{+0.44}$ & TF0.5 & \nodata \\
HD 145333 & 2MASS J16140159-2258462 & $-3.08$ & $-0.86_{-0.26}^{+0.23}$ & $0.44_{-0.27}^{+0.42}$ & $0.31_{-0.19}^{+0.29}$ & $0.46_{-0.28}^{+0.44}$ & TF0.5 & \nodata \\
2MASS J16053815-2039469 & $[$PGZ2001$]$ J160517.9-202420 & $-7.30$ & $-0.26_{-0.28}^{+0.16}$ & $0.33_{-0.18}^{+0.27}$ & $0.31_{-0.17}^{+0.25}$ & $0.40_{-0.22}^{+0.33}$ & TF0.5 & \nodata \\
2MASS J16111742-1918285 & K2-33 & $-1.23$ & $-0.60_{-0.49}^{+0.43}$ & $0.36_{-0.19}^{+0.22}$ & $0.34_{-0.18}^{+0.21}$ & $0.43_{-0.23}^{+0.27}$ & TF0.5 & \nodata \\
2MASS J16051687-1948374 & HD 144273 & $-1.32$ & $-0.58_{-0.78}^{+0.50}$ & $0.33_{-0.12}^{+0.23}$ & $0.29_{-0.11}^{+0.20}$ & $0.20_{-0.08}^{+0.14}$ & TF0.5 & \nodata \\
UCAC4 355-076729 & 2MASS J16033829-1854076 & $-6.00$ & $-0.24_{-0.34}^{+0.15}$ & $0.40_{-0.20}^{+0.32}$ & $0.41_{-0.21}^{+0.33}$ & $0.44_{-0.23}^{+0.35}$ & TF0.5 & \nodata \\
HD 144925 & $[$PGZ2001$]$ J160900.7-190852 & $-2.38$ & $-0.04_{-0.12}^{+0.12}$ & $0.35_{-0.05}^{+0.05}$ & $0.22_{-0.03}^{+0.03}$ & $0.36_{-0.06}^{+0.06}$ & TF0.5 & \nodata \\
\enddata
\tablecomments{The notations in this table are the same as in Table \ref{table:all_scocen_encounters}. $\mathrm{RV}_\text{sim}$ represents the RV value of the flyby star, which do not have RV measurement, that outputs the minimum $d_\text{ca}$.}
\end{deluxetable*}
\end{longrotatetable}

\end{CJK*}
\end{document}